\documentclass[sigconf]{acmart}
\setcopyright{none}
\usepackage{multirow}
\usepackage{listings}
\newfloat{listing}{tbhp}{lol}
\floatname{listing}{Listing}
\usepackage{xcolor}
\definecolor{MyGreen}{HTML}{81B365}
\definecolor{MyPurple}{HTML}{9773A6}
\usepackage{colortbl}
\usepackage{algorithm}
\usepackage{enumitem}
\usepackage{makecell}
\usepackage{pifont}  

\usepackage[most]{tcolorbox}
\definecolor{background}{HTML}{FFFFF5}
\definecolor{edge}{HTML}{87C2B1}
\newtcolorbox{mybox}{colback=background!35,
    colframe=MyGreen!80, 
    width=\columnwidth,
    arc=1mm,
    auto
    outer
    arc
}
\AtBeginDocument{%
  }

\begin{document}

\title{CodeGlance: Understanding Code Reasoning Challenges in LLMs through Multi-Dimensional Feature Analysis}

\author{Yunkun Wang}
\authornote{Both authors contributed equally to this research.}
\affiliation{
  \institution{Zhejiang University}
  \city{Hangzhou}
  \country{China}}
\email{wangykun@zju.edu.cn}

\author{Xuanhe Zhang}
\authornotemark[1]
\affiliation{%
  \institution{Zhejiang University}
  \city{Hangzhou}
  \country{China}}
\email{22560080@zju.edu.cn}

\author{Junxiao Han}
\affiliation{%
 \institution{Hangzhou City University}
 \city{Hangzhou}
 \country{China}}
\email{hanjx@hzcu.edu.cn}

\author{Chen Zhi}
\authornote{Corresponding author}
\affiliation{%
 \institution{Zhejiang University}
 \city{Hangzhou}
 \country{China}}
\email{zjuzhichen@zju.edu.cn}

\author{Shuiguang Deng}
\authornotemark[2]
\affiliation{%
  \institution{Zhejiang University}
  \city{Hangzhou}
  \country{China}}
\email{dengsg@zju.edu.cn}

\renewcommand{\shortauthors}{Wang et al.}

\begin{abstract}
In modern software development, developers frequently need to understand code behavior at a glance---whether reviewing pull requests, debugging issues, or navigating unfamiliar codebases. This ability to reason about dynamic program behavior is fundamental to effective software engineering and increasingly supported by Large Language Models (LLMs). However, existing studies on code reasoning focus primarily on isolated code snippets, overlooking the complexity of real-world scenarios involving external API interactions and unfamiliar functions. This gap hinders our understanding of what truly makes code reasoning challenging for LLMs across diverse programming contexts.

We present \textit{CodeGlance}, a multi-dimensional benchmark investigating code reasoning challenges across three realistic scenarios: intrinsic logic reasoning, API interaction reasoning, and unseen function reasoning. Through systematic evaluation of 7 state-of-the-art LLMs, we reveal that unseen function reasoning poses significant challenges especially for smaller models, with Qwen2.5-3b achieving only 6.0\% accuracy on unseen functions compared to 37.5\% on familiar APIs. We identify critical code complexity features---including execution trace length, API invocation count, and control flow complexity---that significantly impact code reasoning difficulty across scenarios. We further investigate how common augmentation strategies, including CoT, document retrieval, and code search, can improve reasoning performance, finding that their effectiveness varies substantially depending on whether challenges stem from logical complexity or knowledge gaps. These findings provide actionable guidance for developing more capable code reasoning systems and deploying LLM-based programming assistants in real-world software development.
\end{abstract}

\ccsdesc[300]{Software and its engineering~Automatic programming}

\keywords{Code Reasoning, Program Behavior, Evaluation}

\maketitle

\section{Introduction}
In modern software development, developers frequently need to quickly understand code behavior at a glance---whether reviewing pull requests, localizing bugs, detecting vulnerabilities, or refactoring legacy systems. Accurate code comprehension serves as the foundation for these critical downstream tasks. Consider a developer examining a function snippet in an unfamiliar repository: they must not only parse the syntax but also reason about the dynamic behavior---what values intermediate variables will hold, what the function will return given specific inputs, and how external dependencies will affect execution. This ability to reason about code behavior is fundamental to effective software engineering.

Large Language Models (LLMs) have demonstrated remarkable capabilities in code understanding and are now widely deployed in AI-assisted development tools such as GitHub Copilot~\cite{copilot}, Cursor~\cite{cursor}, and various code intelligence platforms. Models like Qwen-Coder~\cite{qwen25}, ChatGPT~\cite{gpt4o}, and DeepSeek~\cite{deepseek-v3} have shown impressive performance on tasks ranging from program completion~\cite{23repocoder,25exploracoder} to bug repair~\cite{24selfrepair,25inspectcoder}. However, a critical question remains: \textit{Can LLMs truly understand code behavior across diverse and realistic coding scenarios?}

While substantial research has explored LLMs' capabilities in static code understanding~\cite{20codesearchnet, 21codexglue,24codescope, 25codesumbeyondfunc, 25designpattern}---such as code summarization, documentation generation, and syntax analysis---their ability to reason about \textit{dynamic program behavior} remains underexplored. Code behavior reasoning differs fundamentally from static explanation: it requires models to simulate program execution, track variable states, understand control flow, and predict concrete outputs given specific inputs. More recently, researchers have begun constructing code reasoning benchmarks to evaluate this capability. 
For example, CruxEval~\cite{cruxeval} assesses LLMs’ performance on the task of input-output prediction for a LLM-generated self-contained function. REval~\cite{reval} evaluated LLM's undertanding for intermediate code behavior such as branch prediction tasks using ClassEval~\cite{23classeval} and HumanEval~\cite{21humaneval}. CodeMind~\cite{codemind} proposed output prediction tasks on existing self-contained code benchmarks. CodeSense~\cite{codesense} collects real world code snippet and evaluate LLM's behavioral understandings.
Despite these efforts, existing benchmarks exhibit two critical limitations. First, most focus primarily on simple, self-contained code snippets that operate in isolation from third-party code libraries, overlooking the complexity inherent in real-world scenarios such as \textbf{external API interactions}---where code behavior emerges from compositional API logic interactions---and \textbf{custom function references}---where critical logic is hidden in unseen functions defined elsewhere in the codebase. Second, existing work provides limited insight into how various scenario-specific \textbf{complexity features impact LLM reasoning performance}, lacking systematic analysis on the challenges posed by expanding factors like execution trace length (whether encapsulated or not), API invocation count, or knowledge requirements across different scenarios.

To address these gaps, we present \textbf{CodeGlance}, a comprehensive benchmark designed to evaluate LLMs' code behavior reasoning capabilities across multi-dimensional coding scenarios. CodeGlance is built on three orthogonal difficulty dimensions that mirror real-world development complexity: (1) \textit{Intrinsic Logic Reasoning}, with 800 self-contained function snippets from CRUXEval~\cite{cruxeval}, testing models' ability to reason about core programming logic without external dependencies; (2) \textit{Common API Behavior Reasoning}, with 1,000+ snippets from DS1000~\cite{ds1000} that depend on popular external libraries (e.g., NumPy, Pandas, PyTorch), evaluating models' understanding of widely-used API semantics within compositional interactions; and (3) \textit{Unseen Function Behavior Reasoning}, with 100 snippets from PanNumEval~\cite{zan_diffcoder_nodate} and MonkBeatEval~\cite{25exploracoder}, where we compare performance on familiar APIs versus novel custom functions to isolate the impact of knowledge distribution shift. We demonstrate the novelty of CodeGlance in Table~\ref{tab:benchmark_features}.
\begin{table}[t]
\centering
\small
\caption{Comparison of code understanding benchmarks. \textit{Std. libs, single}: Simple Python standard libraries (e.g., os, time) with single API call; \textit{3rd-party, multi}: Third-party developed libraries (e.g., Pandas, scipy) with multiple API interactions.}
\resizebox{\columnwidth}{!}{%
\begin{tabular}{lcccc}
\toprule
\textbf{Benchmark} & \textbf{\makecell{Dynamic\\Behavior}} & \textbf{\makecell{Feature\\Analysis}} & \textbf{\makecell{External API\\Dependency}} & \textbf{\makecell{Custom Func\\Interaction}}\\
\midrule
CodeXGlue~\cite{21codexglue} & \textcolor{red}{\ding{55}} & \textcolor{red}{\ding{55}} & \textcolor{red}{\ding{55}} & \textcolor{red}{\ding{55}} \\ 
CruxEval~\cite{cruxeval} & \textcolor{green}{\ding{51}} & \textcolor{red}{\ding{55}} & \textcolor{red}{\ding{55}} & \textcolor{red}{\ding{55}} \\
REval~\cite{reval} & \textcolor{green}{\ding{51}} & \textcolor{red}{\ding{55}} & \textcolor{red}{\ding{55}} & \textcolor{red}{\ding{55}} \\
CodeMind~\cite{codemind} & \textcolor{green}{\ding{51}} & \textcolor{gray}{2 features} & \textcolor{red}{\ding{55}} & \textcolor{red}{\ding{55}} \\
CodeSense~\cite{codesense} & \textcolor{green}{\ding{51}} & \textcolor{red}{\ding{55}} & \textcolor{gray}{Std. libs, single} & \textcolor{red}{\ding{55}} \\
\midrule
\textbf{CodeGlance} & \textcolor{green}{\ding{51}} & \textbf{\textcolor{teal}{9 features}} & \textbf{\textcolor{teal}{3rd-party, multi}} & \textcolor{green}{\ding{51}}\\
\bottomrule
\end{tabular}%
}
\label{tab:benchmark_features}
\vspace{-2em}
\end{table}
Using CodeGlance, we conduct an in-depth empirical study on 7 state-of-the-art LLMs, guided by three core research questions: \textbf{RQ1:} \textit{How do LLMs perform on code behavior reasoning across different programming scenarios?} \textbf{RQ2:} \textit{How do different code features impact LLMs' reasoning performance?} \textbf{RQ3:} \textit{How effective are common enhancement strategies across different code reasoning scenarios?} Our evaluation reveals that unseen function reasoning poses the greatest challenge, particularly for smaller models, dynamic execution features consistently impact reasoning more than static code structure, and enhancement strategies show scenario-dependent effectiveness---CoT excels when knowledge suffices but logic is complex, while knowledge augmentation proves more effective for larger models with genuine knowledge gaps.

In summary, this work makes the following contributions:

\begin{itemize}
    \item \textbf{A comprehensive multi-dimensional benchmark.} We introduce CodeGlance~\cite{codeglance_artifact}, a novel benchmark spanning three orthogonal difficulty levels---intrinsic logic, API interactions, and unknown functions---providing more realistic evaluation than existing single-scenario benchmarks.
    
    \item \textbf{Systematic feature impact analysis.} Through analyzing 9 inherent characteristics in code, we identify which features pose the greatest challenges in different scenarios, offering actionable insights for model improvement.
    
    \item \textbf{Validated augmentation strategies.} We evaluate three optimization approaches (chain-of-thought reasoning, document retrieval, code search) across all scenarios, demonstrating their differential effectiveness and providing practical guidelines for real-world deployment.
    
    \item \textbf{Design implications for code intelligence tools.} Our findings reveal where current LLMs excel and struggle, offering actionable insights for improving AI-integrated programming systems.
\end{itemize}

\section{Benchmark Construction}

\begin{figure*}[htbp]
  \centering
  \includegraphics[width=1\linewidth]{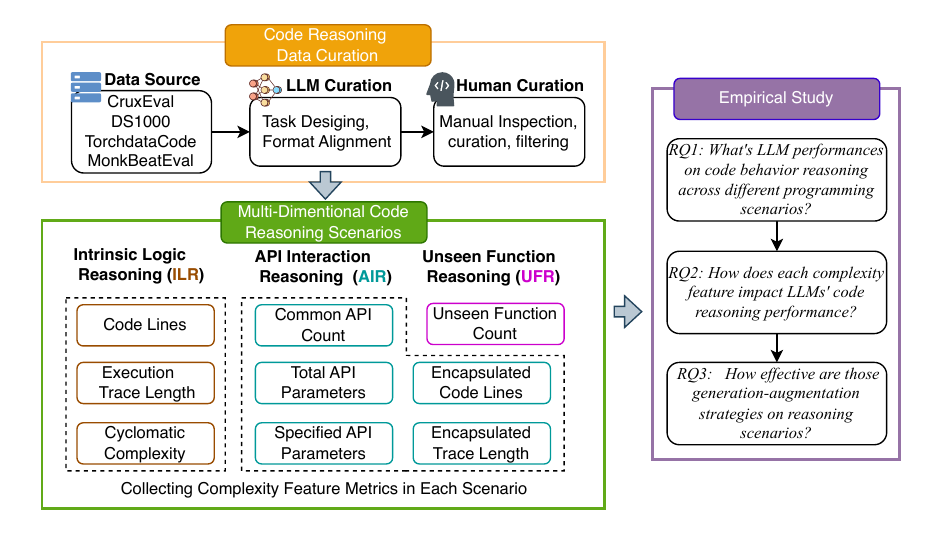}
  \caption{Pipeline of CodeGlance Benchmark Construction and Framework for Empirical Study.}
  \Description{Overview diagram of the CodeGlance pipeline and empirical study framework, showing benchmark construction stages and evaluation workflow from dataset sources through problem curation, feature extraction, and model evaluation.}
  \label{fig:codeglance_framework}
\end{figure*}
\subsection{Task Definition}

In alignment with prior work on code reasoning~\cite{24CruxEval,24CodeMind,ds1000}, we formalize the code behavior reasoning task as follows: Given a code snippet $C$, a test input $I$, and an expression $E$ that captures the program's execution result, the goal is to predict the value of $E$ after executing $C$ with input $I$. 

Formally, let $\mathcal{C}$ denote the space of all possible code snippets, $\mathcal{I}$ denote the space of test inputs, and $\mathcal{V}$ denote the space of possible output values. The code behavior reasoning task can be defined as:

\begin{equation}
f: (\mathcal{C}, \mathcal{I}, \mathcal{E}) \rightarrow \mathcal{V}
\end{equation}

where $f$ represents the reasoning LLM that maps a tuple of code snippet $C \in \mathcal{C}$, test input $I \in \mathcal{I}$, and expression $E \in \mathcal{E}$ to the predicted output value $v \in \mathcal{V}$. The expression $E$ typically takes the form of an assertion statement such as \texttt{assert E(x, y) == <predict>}, where the model must determine the concrete value that makes the assertion true.

\subsection{Identifying a Spectrum of Practical Code Reasoning Scenarios}

To comprehensively evaluate LLMs' code behavior reasoning capabilities and investigate which intrinsic code characteristics hinder their performance, we first identify and define three realistic programming scenarios that developers frequently encounter: \textit{Intrinsic Logic Reasoning}, \textit{API Interaction Reasoning}, and \textit{Unseen Function Reasoning}. For each scenario, we carefully select appropriate data sources that authentically represent the corresponding challenges.

\paragraph{\textbf{Intrinsic Logic Reasoning.}}
In everyday software development, developers routinely examine compact code snippets to understand their behavior—whether during code review, debugging, or refactoring. This scenario represents the fundamental challenge of reasoning about program logic without external dependencies. We aim to evaluate LLMs' ability to reason about program behavior within concentrated code logic with varying degrees of complexity. To isolate the impact of external API interactions, we focus on self-contained code snippets that encapsulate all necessary logic internally. \textbf{CRUXEval}~\cite{cruxeval} is a well-established code reasoning benchmark containing 800 problems with code snippets ranging from 3-13 lines and cyclomatic complexity from 1-5, making it ideal for studying how intrinsic code complexity affects reasoning performance.

\paragraph{\textbf{API Interaction Reasoning.}}
Modern software development heavily relies on invoking well-established APIs from popular libraries to accomplish programming tasks efficiently. However, unlike intrinsic logic reasoning where all code is visible, API interaction scenarios present a unique challenge: part of the program logic is encapsulated within the invoked APIs, making it invisible at a glance. The complexity escalates when a code snippet invokes multiple APIs, each configured with specific parameters—the coordinated behavior of these API interactions can make the overall program behavior difficult to comprehend without deep knowledge of each API's semantics. We aim to evaluate LLMs' ability to precisely understand the semantics of common APIs and their parameters, and to reason about the interactive behavior of multiple APIs working in concert. We adapt two prominent code generation benchmarks into code reasoning datasets: \textbf{DS-1000}~\cite{ds1000} contains 1,000 data science problems spanning seven mainstream Python libraries, making it ideal for studying common API reasoning with familiar libraries; \textbf{TorchdataCode}~\cite{24compositionalapi} includes 50 problems involving the relatively new Torchdata library with 3-8 API invocations per problem, allowing us to examine how API familiarity affects reasoning performance.

\paragraph{\textbf{Unseen Function Reasoning.}}
In real-world development, programmers frequently invoke custom functions defined within their own repositories or utilize APIs from newly released libraries that were unavailable during LLM pre-training. This scenario extends beyond API interaction reasoning by introducing the challenge of reasoning about functions that lie outside the distribution of the model's training data. Developers must infer function behavior from contextual clues such as function names, parameter types, and surrounding code patterns. We aim to evaluate LLMs' ability to reason about out-of-distribution code knowledge by inferring function behavior from context when explicit API documentation or prior exposure is unavailable. To systematically investigate this capability, we leverage two complementary datasets: \textbf{PanNumEval}~\cite{zan_diffcoder_nodate} and \textbf{MonkBeatEval}~\cite{25exploracoder}. PanNumEval is a code generation benchmark containing 50 multi-API problems using extensively-trained Pandas and NumPy libraries, serving as our baseline for familiar API reasoning. MonkBeatEval, developed by \cite{25exploracoder}, systematically transforms all API calls in PanNumEval into privately maintained custom functions (Monkey and BeatNum modules) through bidirectional semantic mappings (e.g., \texttt{pandas.iterrows} $\rightarrow$ \texttt{monkey.traversal}), ensuring functional equivalence while guaranteeing novelty to LLMs. This controlled transformation enables us to isolate the impact of function familiarity: by comparing LLM performance on PanNumEval-R (familiar APIs) versus MonkBeatEval-R (unseen functions), we can precisely measure how knowledge distribution shift affects reasoning performance when all other complexity factors remain constant. We adapt both datasets into code reasoning tasks through our curation pipeline (detailed in Section~\ref{sec:construction}).

\begin{table*}[t]
\centering
\caption{Statistics of CodeGlance benchmark datasets.}
\label{tab:codeglance_stats}
\begin{tabular}{lccccc}
\toprule
\textbf{Dataset} & \textbf{\# Data} & \textbf{Avg. Code} & \textbf{Avg. Cyclomatic} & \textbf{Avg. External} & \textbf{Avg. Unseen} \\
 & \textbf{Sample} & \textbf{Line} & \textbf{Complexity} & \textbf{Function} & \textbf{Knowledge} \\
\midrule
CRUXEval & 800 & 5.47 & 2.14 & 0 & 0 \\
DS-1000-R & 1019 & 5.61 & 1.33 & 3.12 & 0 \\
TorchdataCode-R & 50 & 9.56 & 1.82 & 5.12 & 0 \\
PanNumEval-R & 50 & 5.7 & 1.02 & 4.18 & 0 \\
MonkBeatEval-R & 50 & 5.7 & 1.02 & 4.18 & 4.18 \\
\midrule
\textbf{overall} & \textbf{1969} & \textbf{5.81} & \textbf{1.67} & \textbf{2.52} & \textbf{0.47} \\
\bottomrule
\end{tabular}
\end{table*}

\subsection{Code Reasoning Task Construction}
\label{sec:construction}

While CRUXEval already provides data in the code reasoning format, the code generation benchmarks (DS-1000, TorchdataEval, PanNumEval, and MonkBeatEval) require adaptation. Each code generation problem consists of a task description, reference code solution, and test cases. Our intuitive approach is to leverage the test inputs and reference code to construct reasoning problems where models predict the return value: \texttt{reference\_code(test\_input) == ?}

However, this straightforward conversion faces several challenges: (1) \textbf{Complex return types}: Functions may return nested complex objects (e.g., matplotlib plot objects) that LLMs cannot describe textually, making verification impossible; (2) \textbf{Non-deterministic behavior}: Functions depending on random numbers or system information produce unpredictable results; (3) \textbf{Numerical complexity}: Functions involving intricate numerical computations exceed LLMs' arithmetic capabilities. To address these challenges, we employ a multi-stage construction pipeline combining LLM-based generation with human verification.

\paragraph{Stage 1: LLM-Assisted Problem Generation.}
We prompt an LLM to generate semantically meaningful assertion statements from the reference code and test inputs. Specifically, we instruct the LLM to create assertions that: (1) invoke the reference function with test inputs, (2) extract simple, verifiable properties (strings or integers) from the return value, and (3) ensure semantic meaningfulness to align with realistic code reasoning scenarios. For example, instead of directly comparing complex objects, the LLM generates assertions like \texttt{assert len(result) == ?} or \texttt{assert result['key'] == ?}. We then execute each generated assertion to collect ground-truth values. Assertions that fail execution are immediately discarded, filtering out malformed or incompatible expressions.

\paragraph{Stage 2: Format Standardization.}
The second LLM pass standardizes the generated problems into a consistent format. Each problem is decomposed into two components: (1) the code snippet containing necessary context and function definitions, and (2) the test input and assertion statement. When multiple test inputs or assertions exist for a single reference solution, we split them into separate problems, ensuring each problem has exactly one input-assertion pair. This standardization facilitates uniform evaluation across all scenarios.

\paragraph{Stage 3: Human Verification.}
Human annotators perform final quality control to ensure dataset reliability. We discard problems that: (1) still produce non-deterministic results despite filtering (e.g., due to implicit randomness), (2) fail to execute correctly and cannot be manually repaired, or (3) involve complex numerical computations or large numbers that would unfairly penalize LLMs due to their known arithmetic limitations rather than code reasoning deficiencies. This step ensures that evaluation focuses purely on code behavior understanding rather than numerical precision.

Through this rigorous pipeline, we construct the \textbf{CodeGlance} benchmark, integrating 1,969 high-quality code reasoning problems across three realistic scenarios. Table~\ref{tab:codeglance_stats} presents detailed statistics for each constituent dataset, showing the distribution of code complexity metrics including average code lines, cyclomatic complexity, external API usage, and unseen function counts.

\subsection{Code Complexity Feature Selection}

To systematically investigate what makes code reasoning challenging for LLMs across different scenarios, we extract scenario-specific features that capture the unique complexity dimensions of each programming context. These features enable fine-grained analysis of how different code characteristics impact reasoning performance.

\paragraph{Intrinsic Logic Features.}
In the intrinsic logic reasoning scenario, the primary challenge lies in understanding and mentally simulating complex control flow and data transformations within self-contained code. We extract three key features to quantify this complexity: (1) \textbf{Code Lines}: the total number of lines in the code snippet, reflecting the amount of logic to comprehend; (2) \textbf{Cyclomatic Complexity}: the number of independent execution paths through the code, measuring control flow complexity from conditional branches and loops; (3) \textbf{Execution Lines}: the actual number of lines executed during program runtime with given inputs, capturing the dynamic complexity of the execution path. While code lines measure static size, execution trace length reveals the true computational steps LLMs must mentally simulate.

\paragraph{API Interaction Features.}
In the API interaction reasoning scenario, the core difficulty stems from understanding how multiple APIs coordinate and how their encapsulated logic affects overall program behavior. We extract five features to characterize this interaction complexity: (1) \textbf{API Count}: the number of external API invocations in the code, indicating the breadth of external dependencies; (2) \textbf{Total API Parameters}: the sum of all parameters across all API calls, reflecting the configuration complexity; (3) \textbf{Specified API Parameters}: the number of explicitly provided (non-default) parameters, measuring how much customization is applied to API behavior; (4) \textbf{Encapsulated Code Lines}: the estimated total lines of code hidden within invoked APIs, quantifying the invisible logic; (5) \textbf{Encapsulated Execution Lines}: the estimated execution lines within API implementations, capturing the hidden dynamic complexity. These features help us understand whether reasoning difficulty arises from API quantity, parameter complexity, or the depth of hidden logic.

\paragraph{Knowledge Features.}
In the unseen function reasoning scenario, the fundamental challenge is inferring behavior of functions outside the model's training distribution. We extract one critical feature: \textbf{Unseen Function Count}, which measures the number of novel function invocations that LLMs have not encountered during pre-training. This feature directly quantifies the degree of knowledge distribution shift, allowing us to assess how well LLMs can generalize their reasoning capabilities to unfamiliar code constructs through contextual inference alone.

\section{Evaluation}
\begin{table*}[t]
\centering
\caption{Performance of LLMs on CodeGlance benchmark across different coding scenarios. Pass@1 and Pass@3 represent the percentage of problems solved with 1 and 3 attempts respectively. We split the CodeGlance benchmark into three scenarios: ILR: Intrinsic Logic Reasoning; AIR: API Interaction Reasoning; UFR: Unseen Function Reasoning, featuring 4 datasets: CX: CRUXEval; DS: DS-1000-R; TD: TorchdataCode-R; MB: MonkBeatEval-R.}
\label{tab:codeglance_main}
\resizebox{\textwidth}{!}{
\begin{tabular}{lcccccccc}
\toprule
\multirow{2}{*}{\textbf{Model}} & \multicolumn{2}{c}{\textbf{ILR-CX}} & \multicolumn{2}{c}{\textbf{AIR-DS}} & \multicolumn{2}{c}{\textbf{AIR-TD}} & \multicolumn{2}{c}{\textbf{UFR-MB}} \\
\cmidrule(lr){2-3} \cmidrule(lr){4-5} \cmidrule(lr){6-7} \cmidrule(lr){8-9}
& \textbf{pass@1} & \textbf{pass@3} & \textbf{pass@1} & \textbf{pass@3} & \textbf{pass@1} & \textbf{pass@3} & \textbf{pass@1} & \textbf{pass@3} \\
\midrule
Qwen2.5-3b & 21.5\% & 38.3\% & 37.5\% & 54.7\% & 45.0\% & 60.5\% & 6.0\% & 14.5\% \\
Qwen2.5-7b & 46.2\% & 56.8\% & 49.9\% & 63.0\% & 65.0\% & 76.5\% & 40.5\% & 57.5\% \\
Qwen2.5-14b & 58.4\% & 59.0\% & 62.0\% & 63.0\% & 68.0\% & 68.0\% & 49.5\% & 50.0\% \\
Qwen2.5-32b & 65.3\% & 68.5\% & 66.2\% & 69.7\% & 68.5\% & 71.5\% & 64.0\% & 79.5\% \\
DeepSeek-Coder-6.7b & 40.4\% & 49.0\% & 44.8\% & 56.4\% & 56.0\% & 60.2\% & 36.2\% & 50.3\% \\
DeepSeek-Coder-7b & 42.5\% & 51.1\% & 47.9\% & 60.2\% & 37.4\% & 60.5\% & 44.8\% & 59.3\% \\
\midrule
GPT-4o-mini & 56.6\% & 59.9\% & 61.1\% & 66.9\% & 66.2\% & 68.0\% & 52.2\% & 58.2\% \\
GPT-4o & 64.2\% & 73.5\% & 73.2\% & 77.7\% & 64.8\% & 69.6\% & 56.0\% & 66.3\% \\
\bottomrule
\end{tabular}
}
\end{table*}

\subsection{Experimental Setup}

\paragraph{Research Questions.}
Our empirical study is guided by three research questions:
\begin{itemize}[leftmargin=*,nosep]
    \item \textbf{RQ1:} How do LLMs perform on code behavior reasoning across different programming scenarios (intrinsic logic, API interaction, and unseen function reasoning)?
    \item \textbf{RQ2:} How do different code features (logic complexity, API characteristics, knowledge requirements) impact LLMs' reasoning performance?
    \item \textbf{RQ3:} How effective are common enhancement strategies (CoT, RAG, code search) across scenarios with varying complexity and knowledge demands?
\end{itemize}

\paragraph{Evaluated Models.}
We evaluate 7 state-of-the-art LLMs representing diverse model families and scales: \textbf{Qwen2.5-Coder}~\cite{qwen25} (3B, 7B, 14B, 32B-Instruct), a family of code-specialized models enabling systematic study of scale effects; \textbf{GPT-4o}~\cite{gpt4o}, a leading closed-source model with strong general-purpose capabilities; \textbf{DeepSeek-Coder-7B-Instruct-v2}~\cite{deepseek-v3}, a competitive open-source code model; and \textbf{Qwen2.5-32B-Instruct}, a general-purpose model for comparison with its code-specialized counterpart. All models are evaluated in their instruction-tuned variants to ensure fair comparison under the same prompting paradigm.

\paragraph{Benchmark and Metrics.}
Our evaluation leverages CodeGlance, comprising 1,969 code reasoning problems across three scenarios: 800 problems from CRUXEval~\cite{cruxeval} for intrinsic logic reasoning, 1,019 problems from DS-1000~\cite{ds1000} and 50 from TorchdataCode~\cite{24compositionalapi} for API interaction reasoning, and 100 problems from PanNumEval/MonkBeatEval~\cite{25exploracoder} for unseen function reasoning. We report \textbf{pass@1} (percentage of problems solved with a single attempt) as the primary metric for measuring reasoning accuracy, and \textbf{pass@3} (success rate with three attempts) to assess models' ability to explore multiple reasoning paths. For feature impact analysis (RQ2), we compute Pearson correlation coefficients and visualize performance trends across feature value ranges. For enhancement strategy evaluation (RQ3), we measure absolute performance improvements relative to direct prompting baselines.

\paragraph{Implementation Details.}
All experiments use greedy decoding (temperature=0) for pass@1 evaluation to ensure reproducibility, and sampling with temperature=0.8 for pass@3 evaluation to encourage diversity. We set max output tokens to 2048 and use a unified prompt template across all models, providing the code snippet and test input while asking models to predict the assertion result. For CoT experiments, we prepend "Let's think step by step" (zero-shot) or provide 2-3 worked examples (few-shot). For RAG experiments, we retrieve relevant API documentation or function source code and include them in the prompt context. All evaluations are conducted on the same hardware configuration, and results are averaged over the complete test set without data filtering.

This section presents the experimental results and analysis for each research question.

\subsection{RQ1: How do LLMs perform on code behavior reasoning across different programming scenarios?}

We evaluate 7 state-of-the-art LLMs across CodeGlance's three orthogonal difficulty scenarios: Intrinsic Logic Reasoning (ILR-CX), API Interaction Reasoning (AIR-DS and AIR-TD), and Unseen Function Reasoning (UFR-MB). Table~\ref{tab:codeglance_main} presents the pass@1 and pass@3 performance across all models and scenarios. 

\paragraph{Model Parameter Scaling Effects.}
Model size strongly correlates with reasoning performance across all scenarios, showing an average 3$\times$ improvement from 3B to 32B parameters within the Qwen2.5-Coder family, though the magnitude of improvement varies substantially.
For intrinsic logic reasoning and API interaction reasoning, we observe a 3.0$\times$ and 1.4$\times$ gains; and for unseen function reasoning, the improvement is a striking 10.7$\times$.
The disproportionate benefit for out-of-distribution reasoning is noteworthy: while the 3B model nearly collapses on unseen functions (6.0\%), the 32B model achieves performance comparable to familiar scenarios (64.0\% vs 65.3\% for ILR, 66.2\% for AIR). This suggests it has developed stronger capabilities to reason about unknown function behavior from surrounding code context, function naming conventions, and parameter usage patterns. 

\paragraph{Performance Patterns Across Scenarios.}
A clear difficulty gradient emerges in smaller models: narrow-domain API reasoning (TorchdataCode) outperforms diverse API composition (DS-1000) and intrinsic logic (CRUXEval), while unseen function reasoning (MonkBeatEval) consistently yields the lowest performance. 
Despite these differences, pass@1 across all scenarios converges to a performance plateaus of 64.0-68.5\% at 32B scale.
This convergence is particularly notable when contrasted with the substantial performance disparities observed at smaller scales.
This may suggest that smaller models struggle with scenario-specific challenges---understanding API semantics, inferring novel functions, tracking control flow---creating large cross-scenario variation. As model capacity increases, these barriers could be progressively overcome, with the gap narrowing from 39 to 4.5 percentage.

\paragraph{Model-Specific Strengths.}
Different models exhibit distinct capabilities: GPT-4o excels in multi-API scenarios (DS-1000: 73.2\%) but underperforms in unseen functions compared to Qwen2.5-32B (56.0\% vs. 64.0\%); DeepSeek-Coder variants also show different relative strengths across API domains.
These patterns suggest no single model consistently outperforms across all scenarios, with relative strengths varying by pretraining data and code characteristics. This motivates ensemble approaches or model routing strategies that select specialized models based on scenario requirements.

\begin{mybox}
  \small
  \textbf{Answer to RQ1:}
  LLMs perform well on intrinsic logic and diverse API reasoning but struggle with unseen functions on smaller models, exposing critical capability gaps in out-of-distribution scenarios. Scaling model parameters substantially improves performance on unseen function reasoning tasks. However, a shared performance ceiling around 65\% indicates persistent difficulties in fundamental code reasoning.
\end{mybox}

\subsection{RQ2: How do different code features impact LLMs' reasoning performance?}

Having established the overall performance landscape in RQ1, we now investigate the underlying factors that drive these results. We systematically analyze 9 code characteristics across three feature categories: (1) \textit{logic features} (code lines, cyclomatic complexity, execution trace length) that capture intrinsic computational complexity; (2) \textit{API features} (API count, parameter count) that quantify external dependency complexity; and (3) \textit{knowledge features} (unseen function count) that measure distribution shift from training data. By examining how these features correlate with reasoning performance across different scenarios, we aim to identify which code properties pose the greatest challenges for LLMs and why.

\subsubsection{Scenario 1: Intrinsic logic reasoning}

\begin{figure}[htbp]
  \caption{Code lines analysis in CRUXEval.}
  \Description{Two-panel figure for CRUXEval code lines.}
  \centering
  \begin{minipage}[b]{0.49\linewidth}
    \centering
    \includegraphics[width=\linewidth]{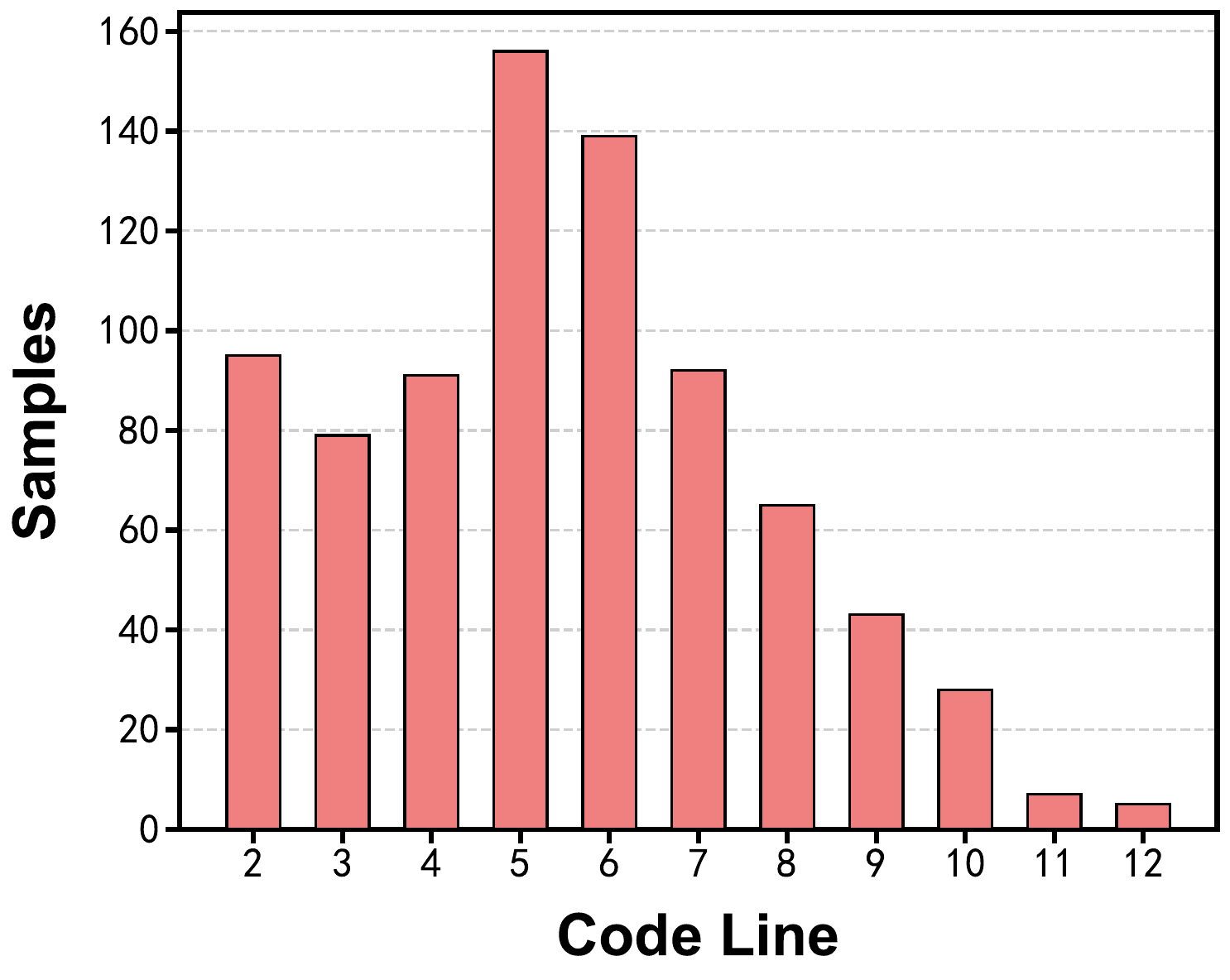}
    \caption*{(a) Distribution}
  \end{minipage}
  \hfill
  \begin{minipage}[b]{0.49\linewidth}
    \centering
    \includegraphics[width=\linewidth]{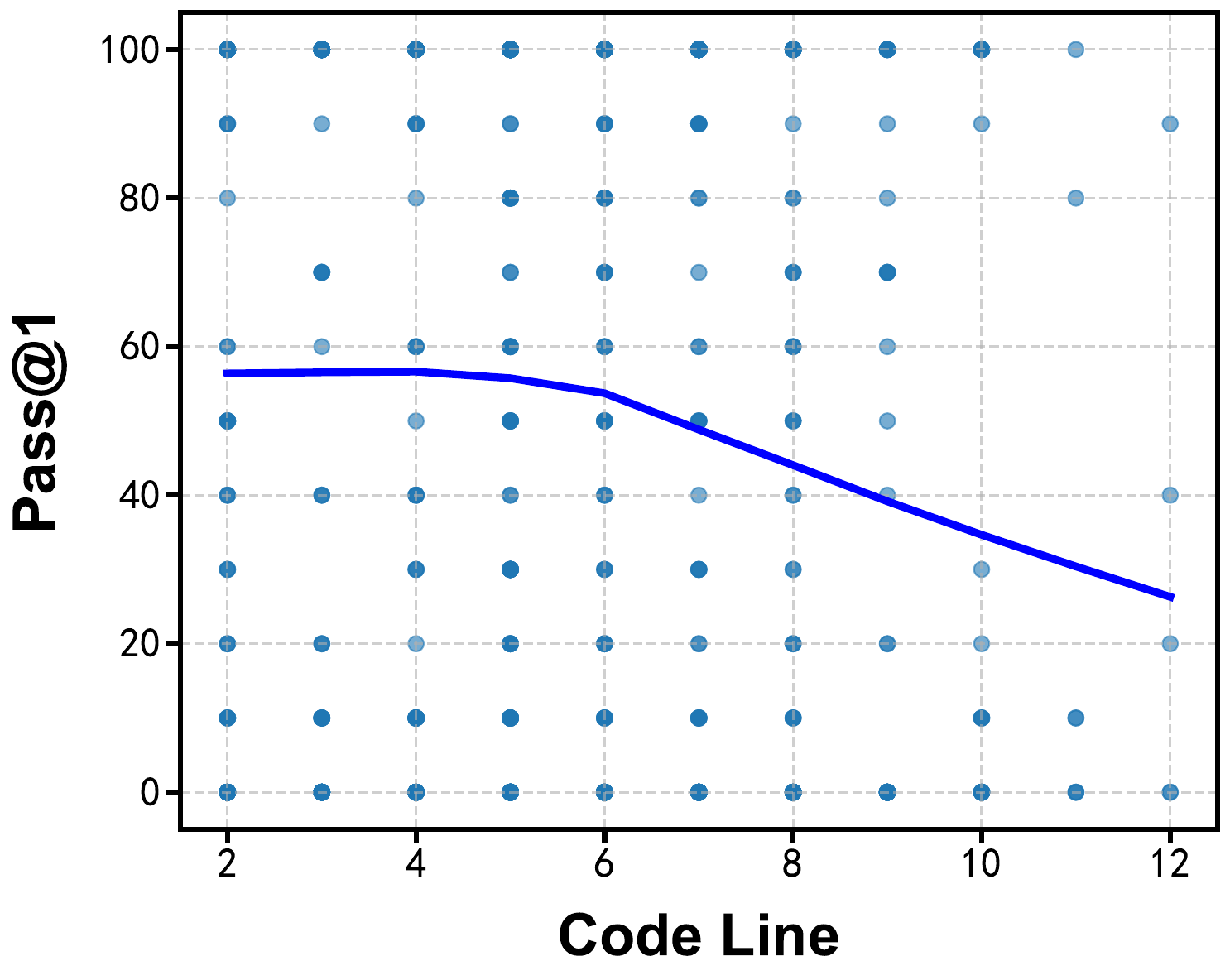}
    \caption*{(b) Performance impact}
  \end{minipage}
  \label{fig:cruxeval_codeline}
\end{figure}

\begin{figure}[htbp]
  \caption{Cyclomatic complexity analysis in CRUXEval.}
  \Description{Two-panel figure for CRUXEval cyclomatic complexity.}
  \centering
  \begin{minipage}[b]{0.49\linewidth}
    \centering
    \includegraphics[width=\linewidth]{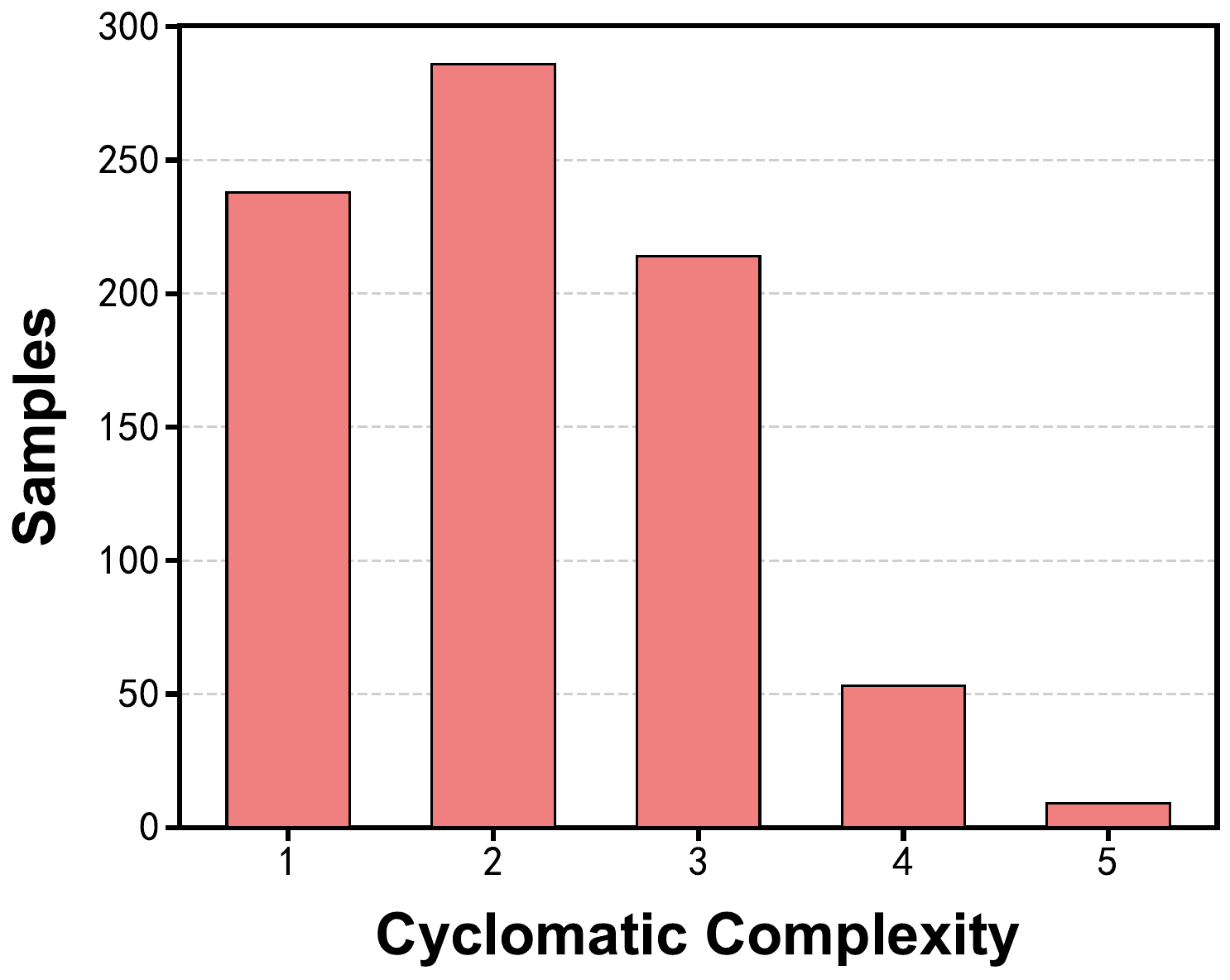}
    \caption*{(a) Distribution}
  \end{minipage}
  \hfill
  \begin{minipage}[b]{0.49\linewidth}
    \centering
    \includegraphics[width=\linewidth]{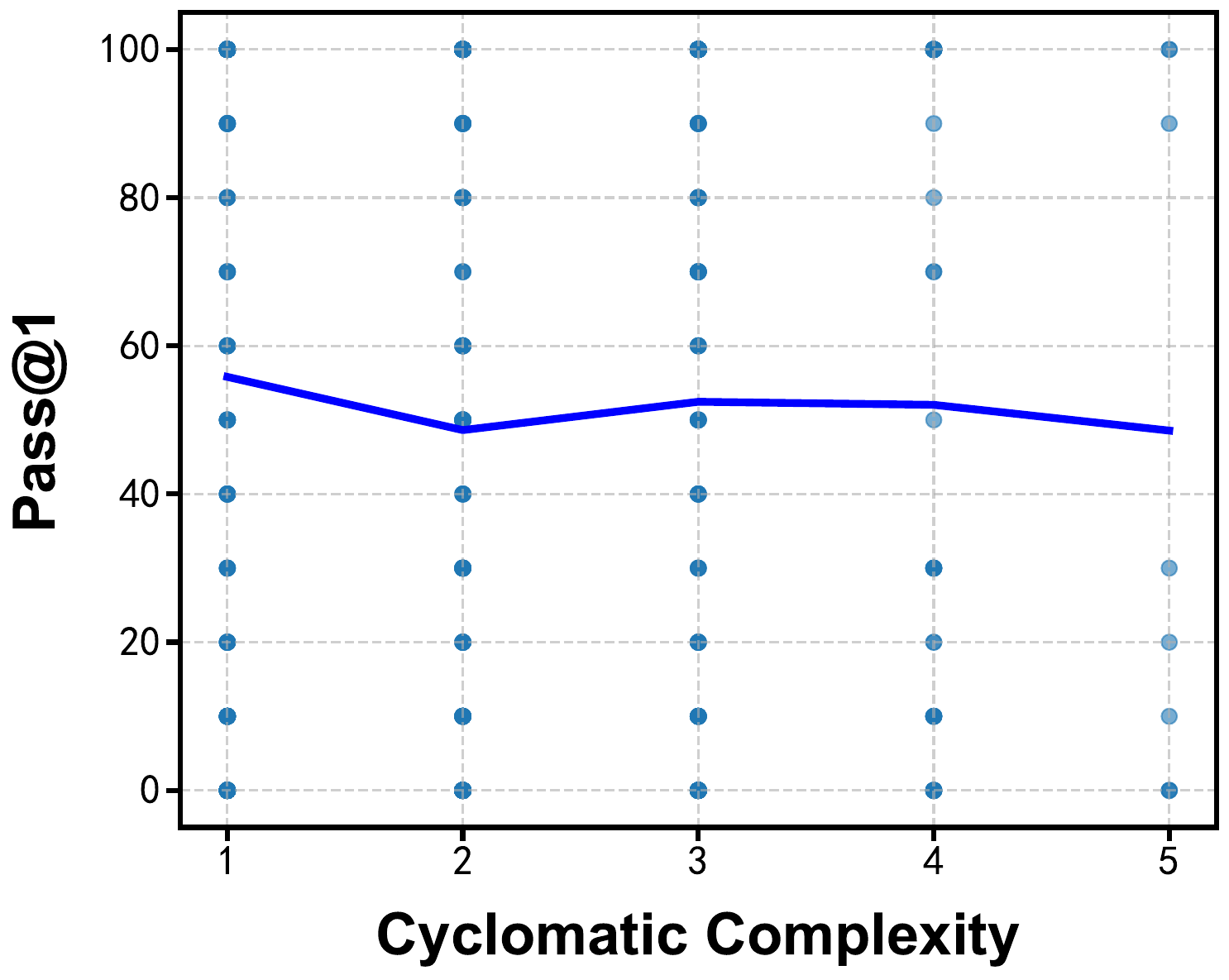}
    \caption*{(b) Performance impact}
  \end{minipage}
  \label{fig:cruxeval_cyclo}
\end{figure}

\begin{figure}[htbp]
  \caption{Execution trace length analysis in CRUXEval.}
  \Description{Two-panel figure for CRUXEval execution trace length.}
  \centering
  \begin{minipage}[b]{0.49\linewidth}
    \centering
    \includegraphics[width=\linewidth,height=0.8\linewidth]{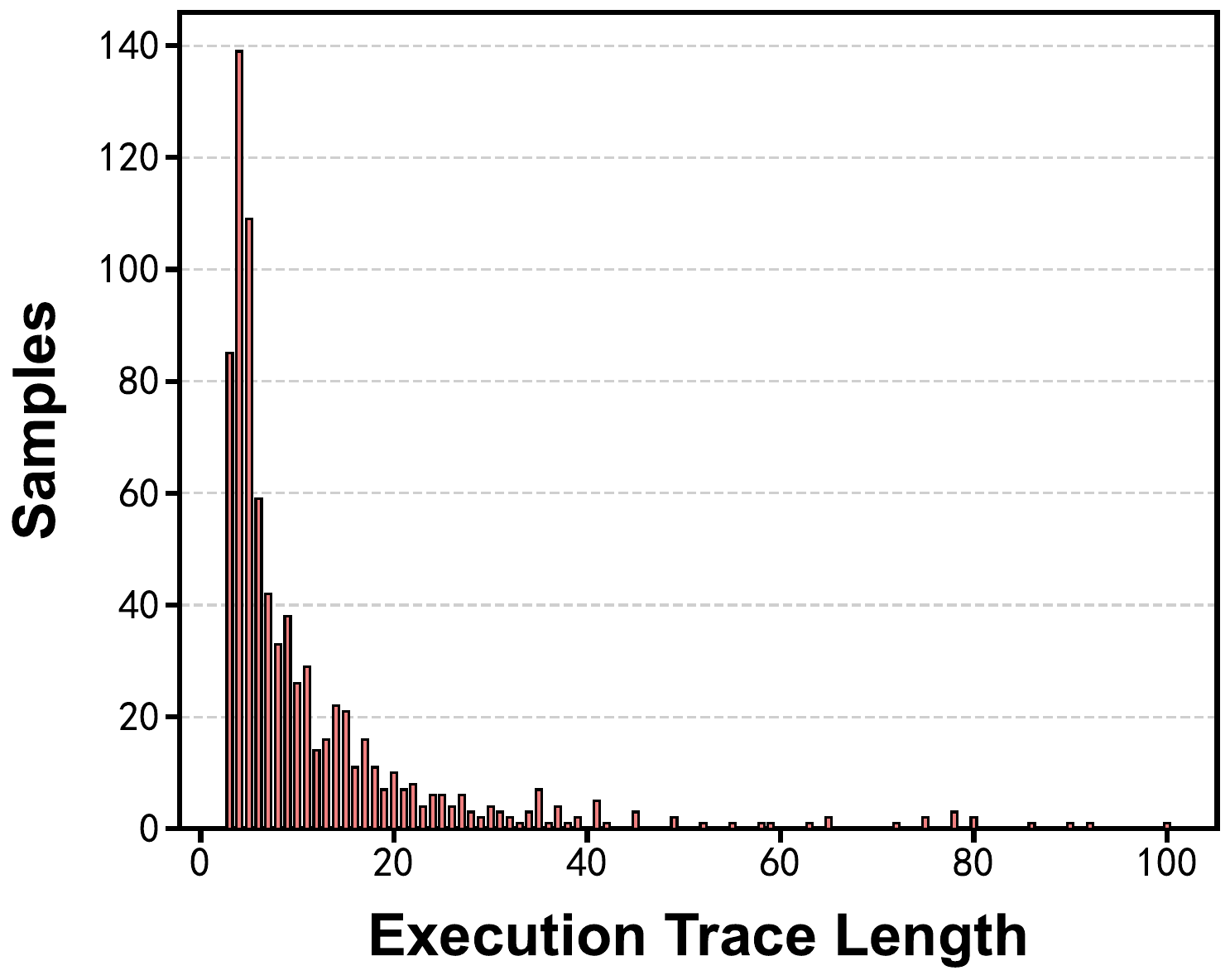}
    \caption*{(a) Distribution}
  \end{minipage}
  \hfill
  \begin{minipage}[b]{0.49\linewidth}
    \centering
    \includegraphics[width=\linewidth]{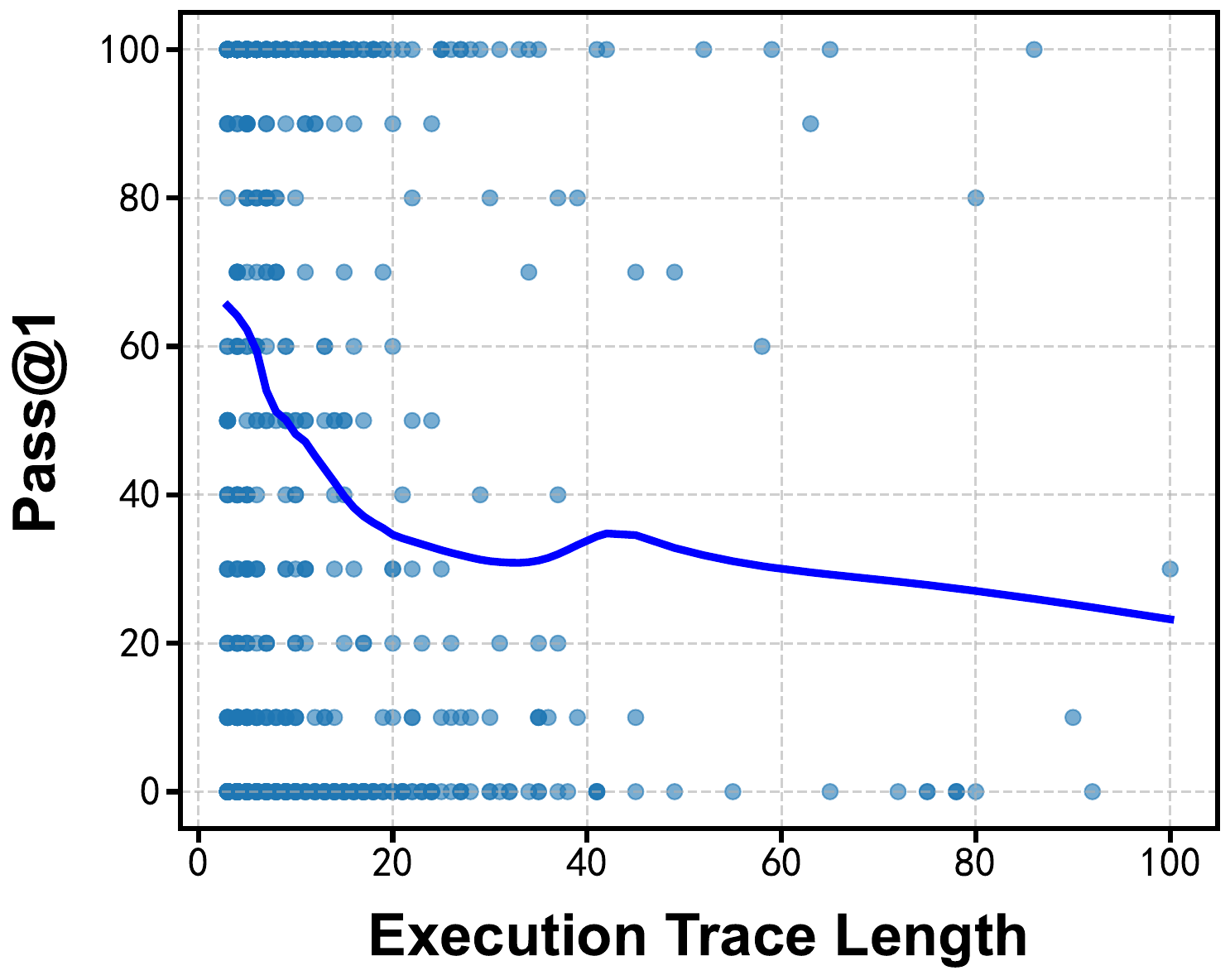}
    \caption*{(b) Performance impact}
  \end{minipage}
  \label{fig:cruxeval_exec}
\end{figure}

We analyze three key features on CRUXEval to understand what makes intrinsic logic reasoning challenging: code lines, cyclomatic complexity, and execution trace length (Figures~\ref{fig:cruxeval_codeline}-\ref{fig:cruxeval_exec}). The bar chart shows the distribution of features in dataset samples. Each blue point in the scatter plot represents a evaluated code sample, with the x-axis indicating the feature value and the y-axis showing the model's normalized pass@1 within 10 tries on that sample. The blue fitting curve indicates the trend of model's code reasoning performance as the feature value increases.

Among the three features, \textbf{execution trace length} emerges as the dominant factor. Performance drops sharply for 30+\% within 20 steps, then continues this degration for longer executions
\textbf{Code length} shows a more gradual decline: performance remains stable for short snippets (2-5 lines) before degrading to approximately 27\% for longer code (12+ lines), indicating that while more static code line does increase difficulty, the effect is less pronounced than the actual executed lines. 
In contrast, \textbf{cyclomatic complexity} exhibits minimal impact---the fitting curve remains relatively flat across all complexity levels (1-5), suggesting that the number of control flow branches alone does not substantially affect reasoning difficulty. This pattern indicates that LLMs can handle branching logic reasonably well when each path remains short; the challenge arises not from branch \textit{count} but from the cumulative execution steps required to trace through those branches. 
Collectively, these findings suggest that augmentation strategies should prioritize decomposing long execution sequences into manageable steps (e.g., intermediate state tracking, step-by-step CoT) rather than simplifying static code structure.

\subsubsection{Scenario 2: Common API Interaction reasoning}

\begin{figure}[htbp]
  \caption{API count analysis in DS-1000.}
  \Description{Two-panel figure for DS-1000 API count.}
  \centering
  \begin{minipage}[b]{0.49\linewidth}
    \centering
    \includegraphics[width=\linewidth]{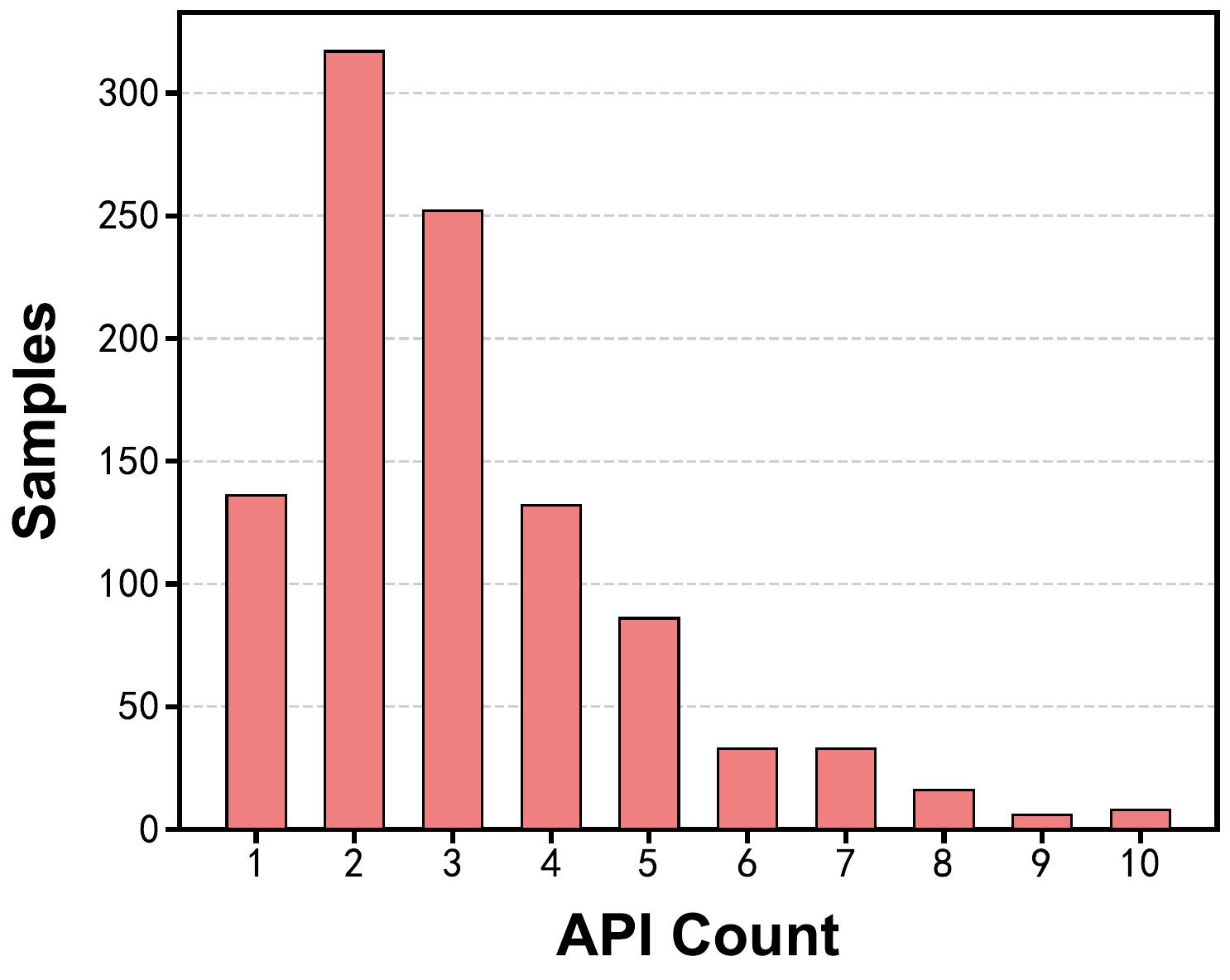}
    \caption*{(a) Distribution}
  \end{minipage}
  \hfill
  \begin{minipage}[b]{0.49\linewidth}
    \centering
    \includegraphics[width=\linewidth]{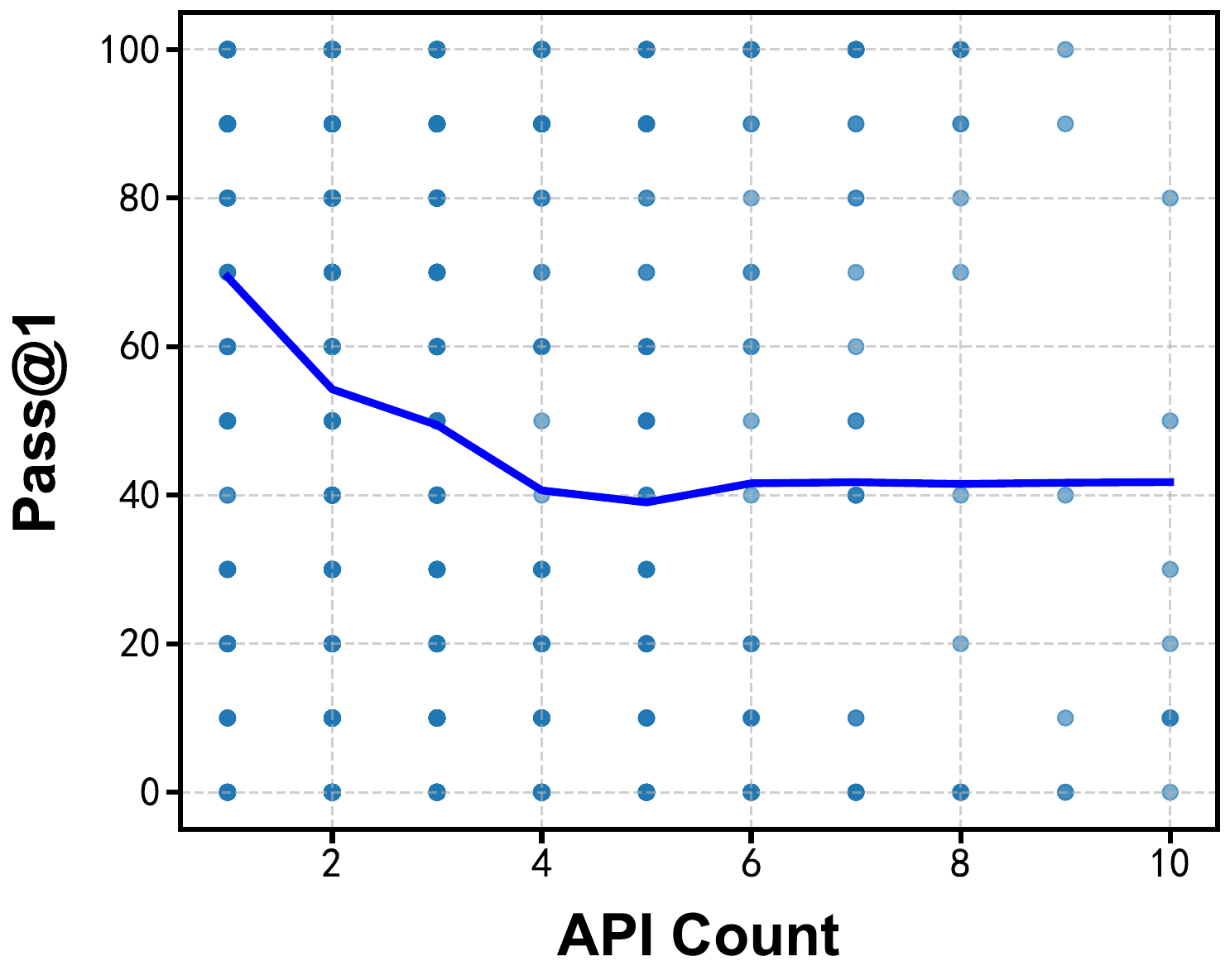}
    \caption*{(b) Performance impact}
  \end{minipage}
  \label{fig:ds1000_apicount}
\end{figure}

\begin{figure}[htbp]
  \caption{Code lines analysis in DS-1000.}
  \Description{Two-panel figure for DS-1000 code lines.}
  \centering
  \begin{minipage}[b]{0.49\linewidth}
    \centering
    \includegraphics[width=\linewidth]{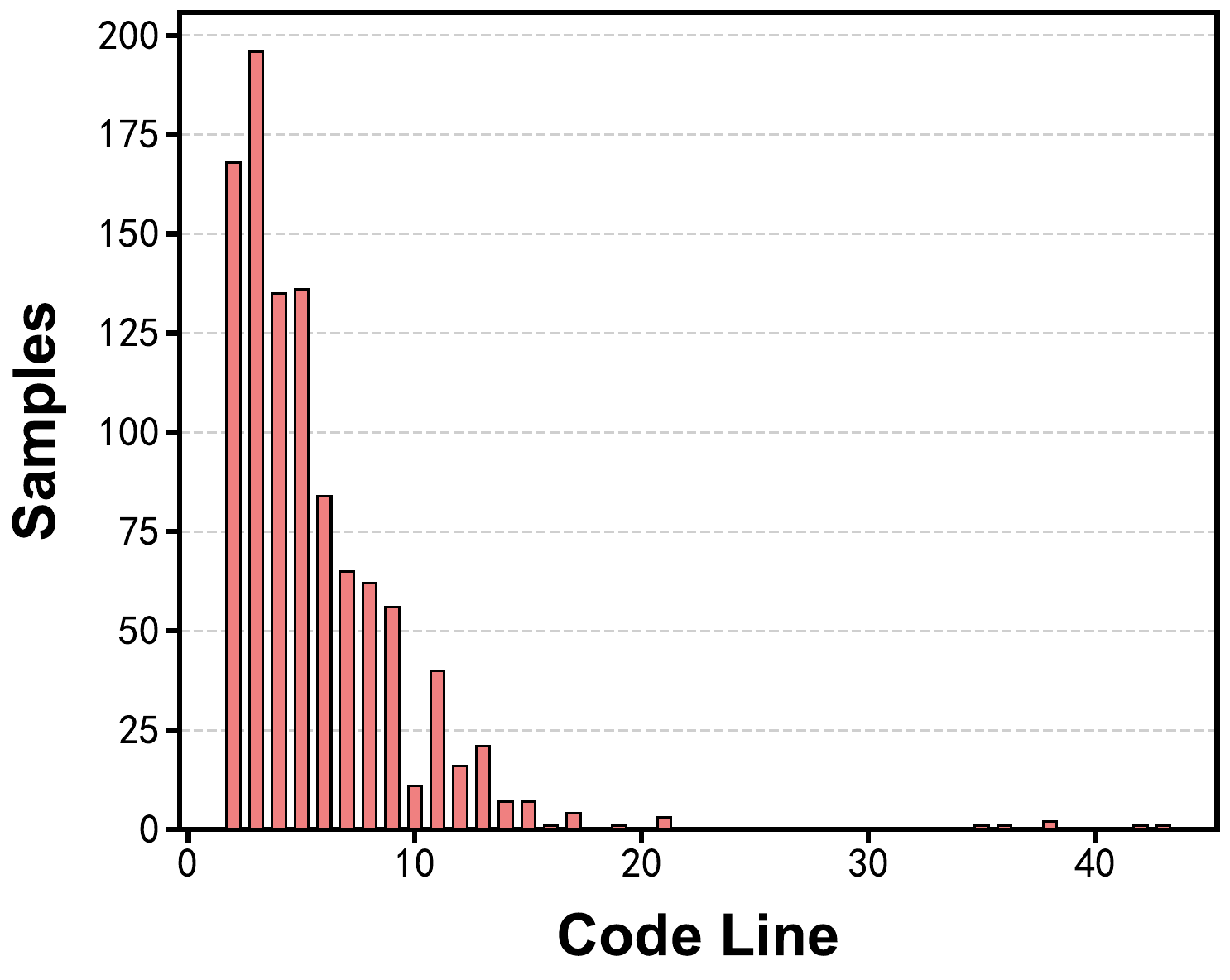}
    \caption*{(a) Distribution}
  \end{minipage}
  \hfill
  \begin{minipage}[b]{0.49\linewidth}
    \centering
    \includegraphics[width=\linewidth]{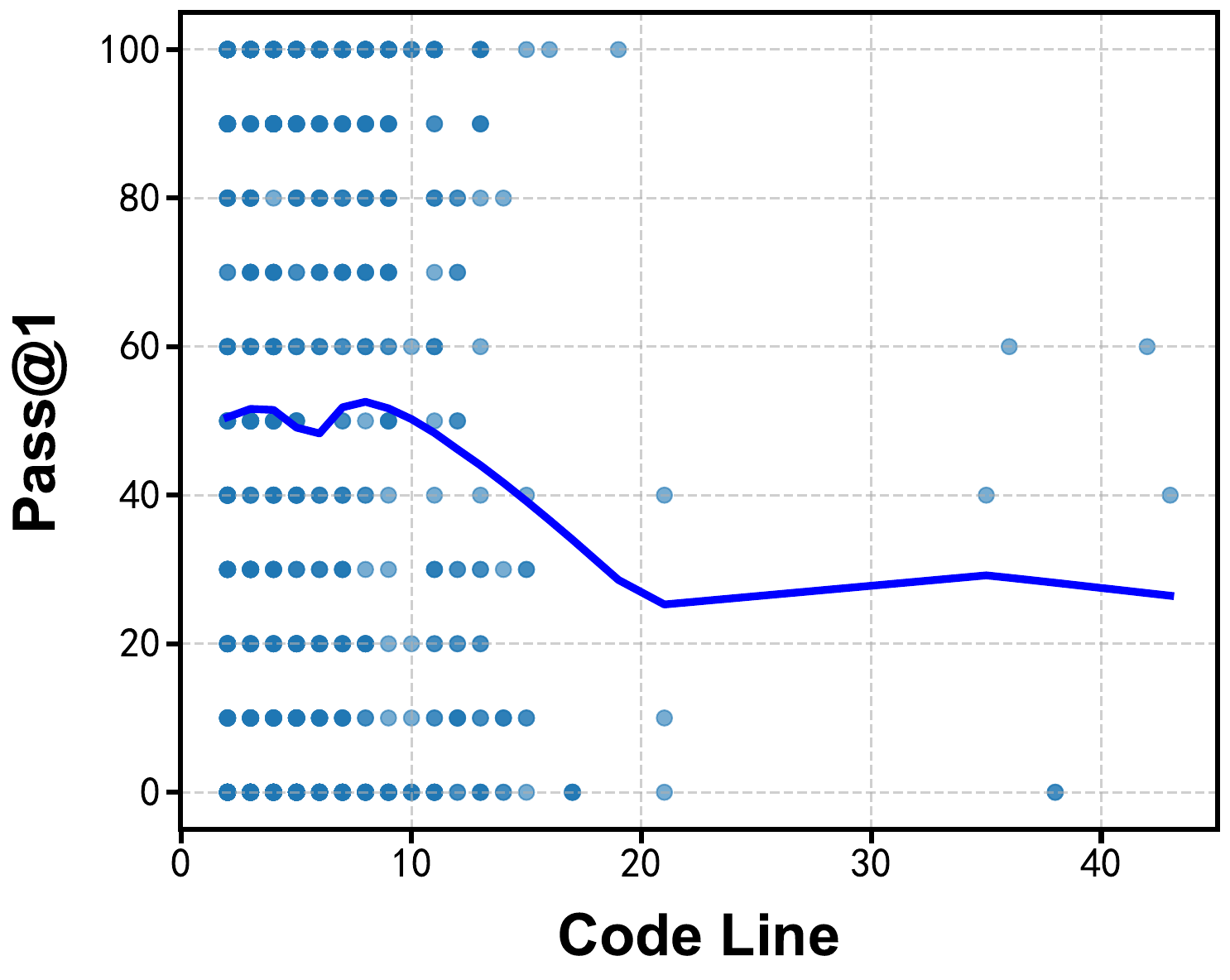}
    \caption*{(b) Performance impact}
  \end{minipage}
  \label{fig:ds1000_codeline}
\end{figure}

\begin{figure}[htbp]
  \caption{Defined parameters analysis in DS-1000.}
  \Description{Two-panel figure for DS-1000 defined parameters.}
  \centering
  \begin{minipage}[b]{0.49\linewidth}
    \centering
    \includegraphics[width=\linewidth]{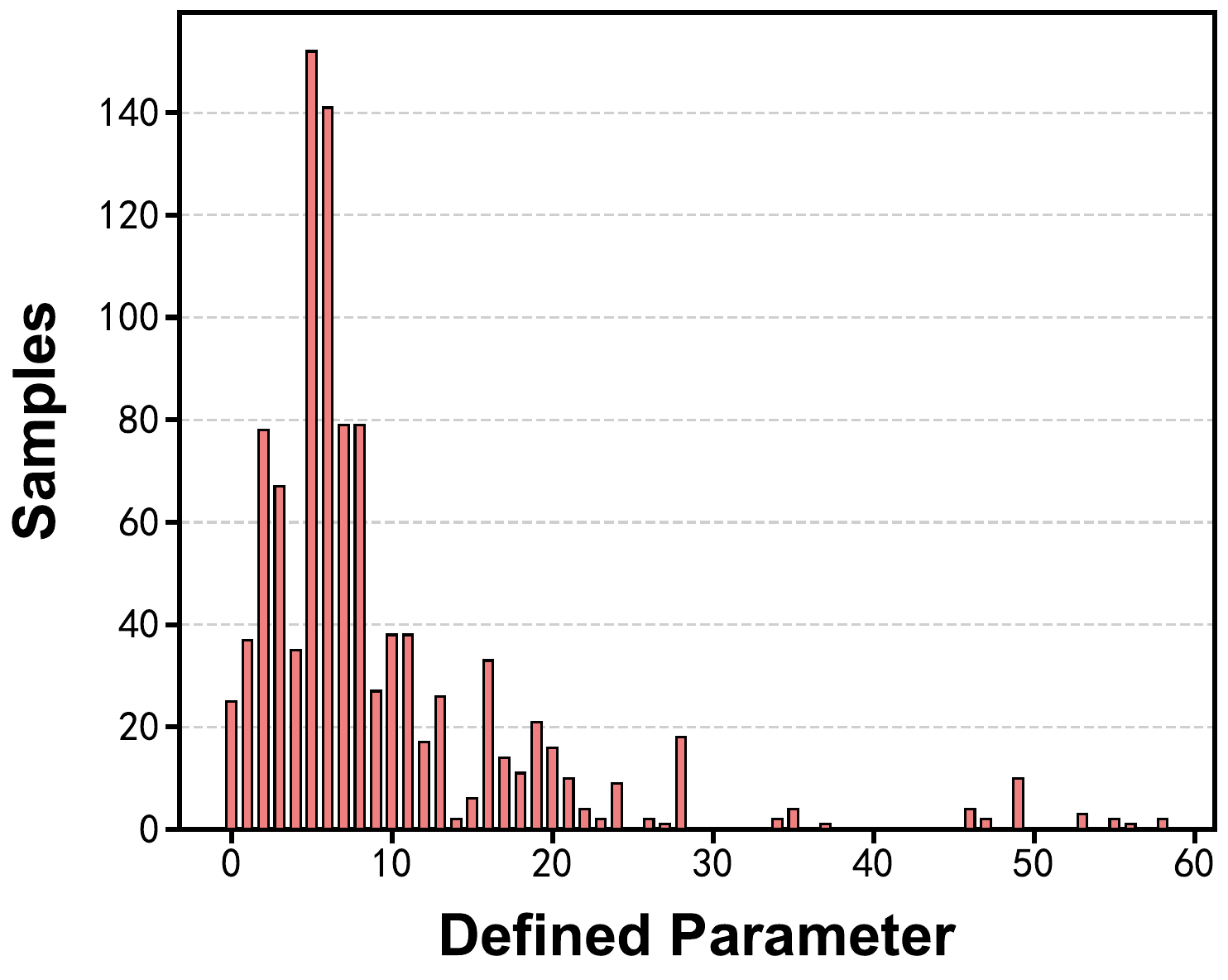}
    \caption*{(a) Distribution}
  \end{minipage}
  \hfill
  \begin{minipage}[b]{0.49\linewidth}
    \centering
    \includegraphics[width=\linewidth]{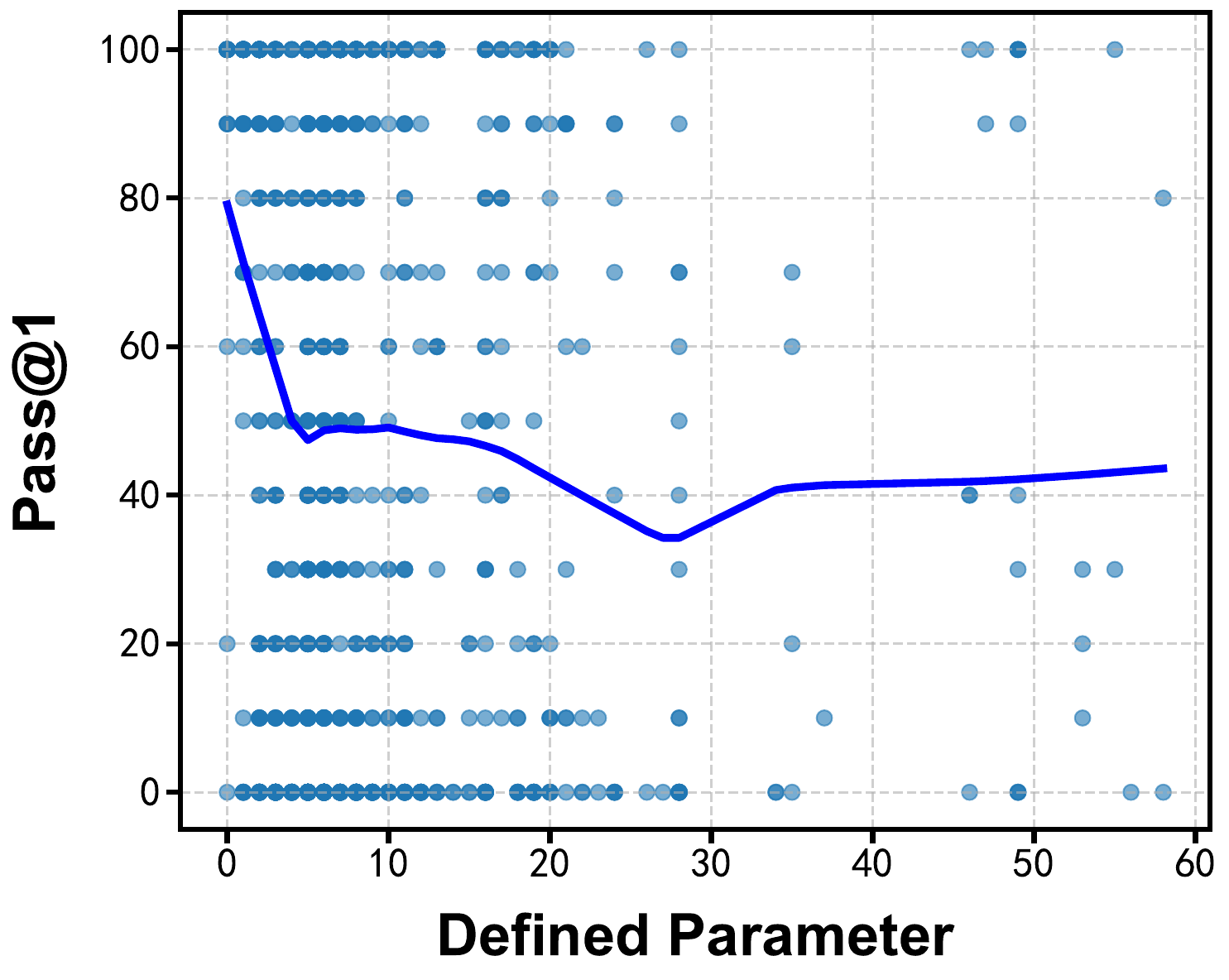}
    \caption*{(b) Performance impact}
  \end{minipage}
  \label{fig:ds1000_defparam}
\end{figure}

\begin{figure}[htbp]
  \caption{Specified parameters analysis in DS-1000.}
  \Description{Two-panel figure for DS-1000 specified parameters.}
  \centering
  \begin{minipage}[b]{0.49\linewidth}
    \centering
    \includegraphics[width=\linewidth]{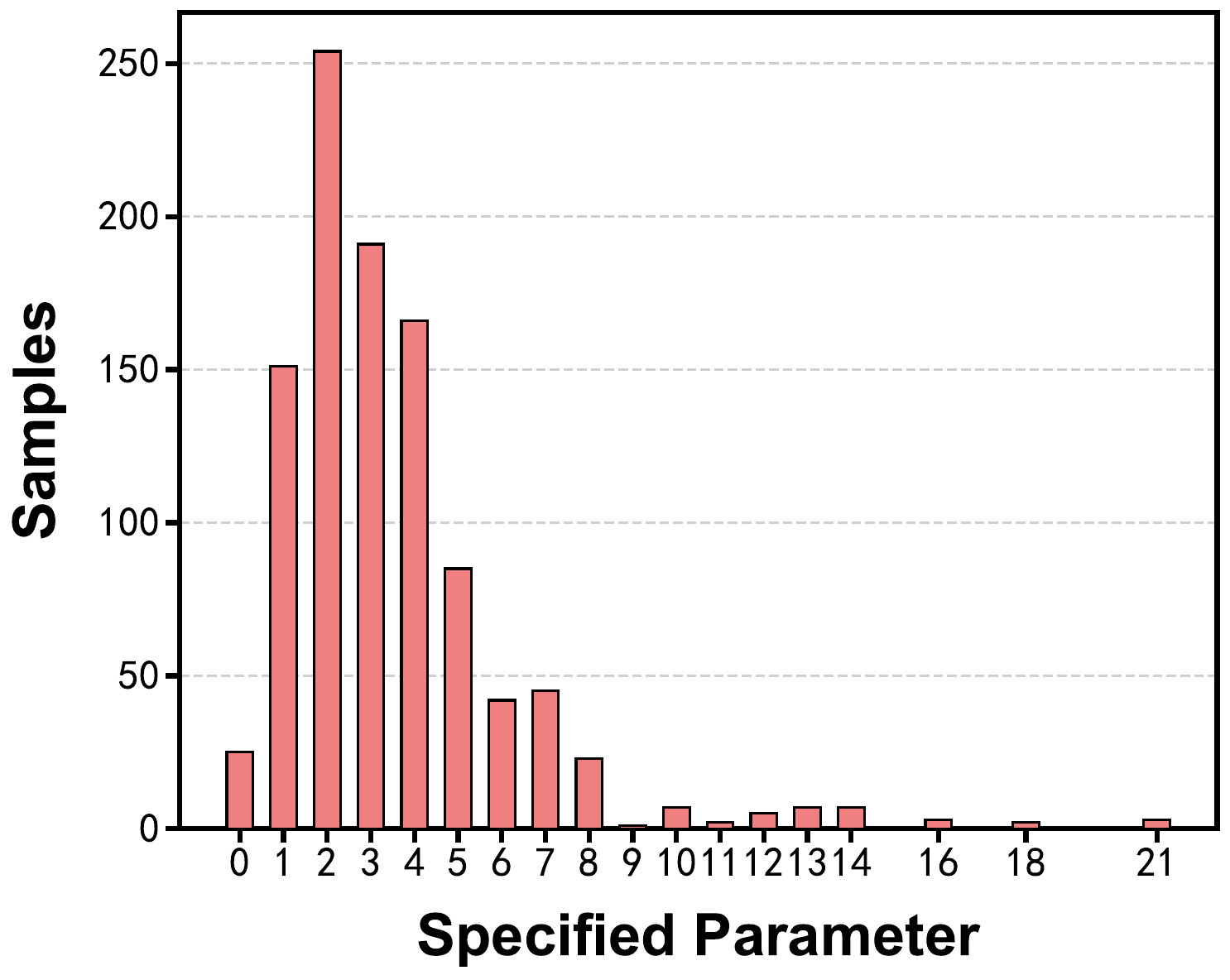}
    \caption*{(a) Distribution}
  \end{minipage}
  \hfill
  \begin{minipage}[b]{0.49\linewidth}
    \centering
    \includegraphics[width=\linewidth]{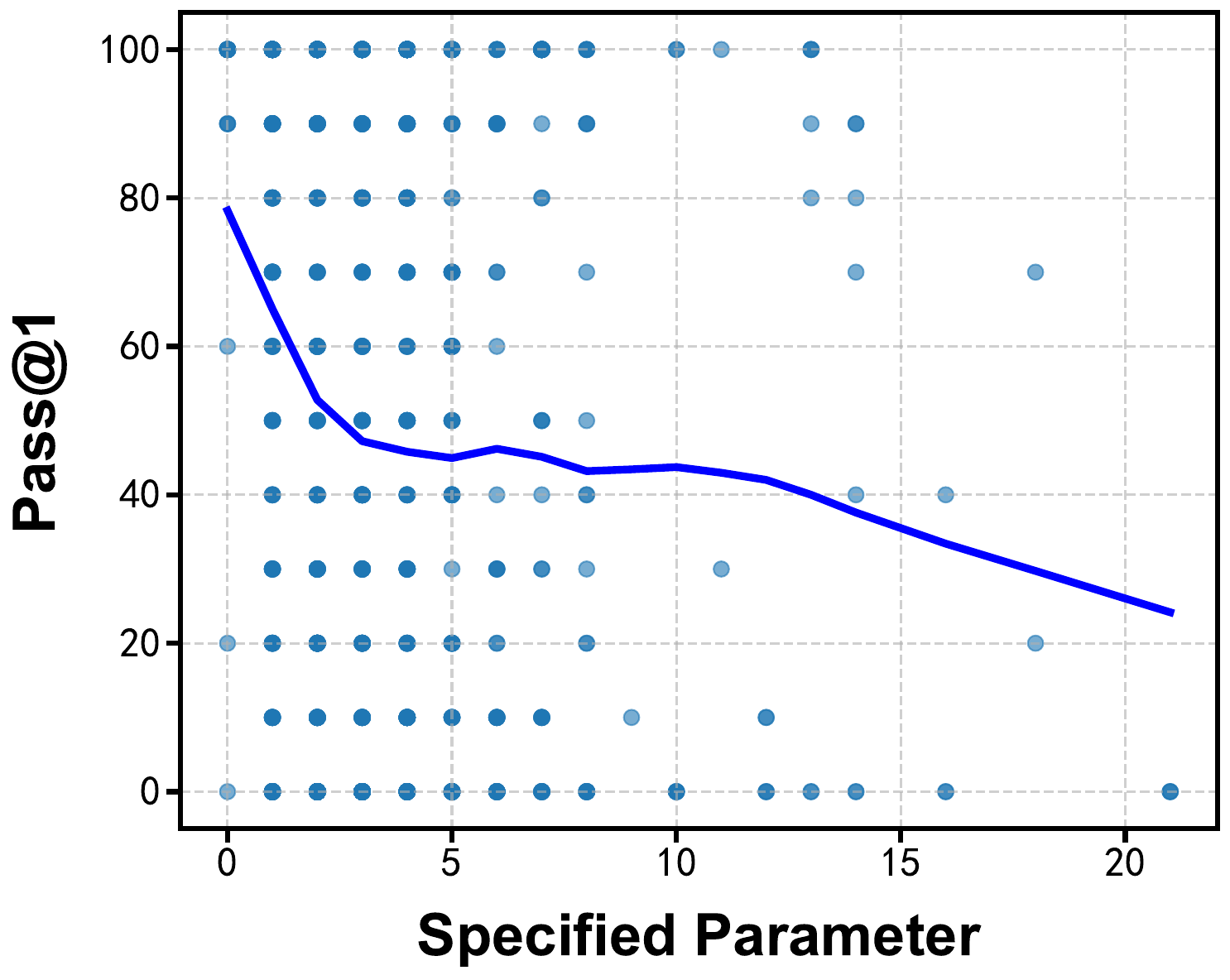}
    \caption*{(b) Performance impact}
  \end{minipage}
  \label{fig:ds1000_execparam}
\end{figure}

\begin{figure}[htbp]
  \caption{Execution trace length analysis in DS-1000.}
  \Description{Two-panel figure for DS-1000 execution trace length.}
  \centering
  \begin{minipage}[b]{0.49\linewidth}
    \centering
    \includegraphics[width=\linewidth]{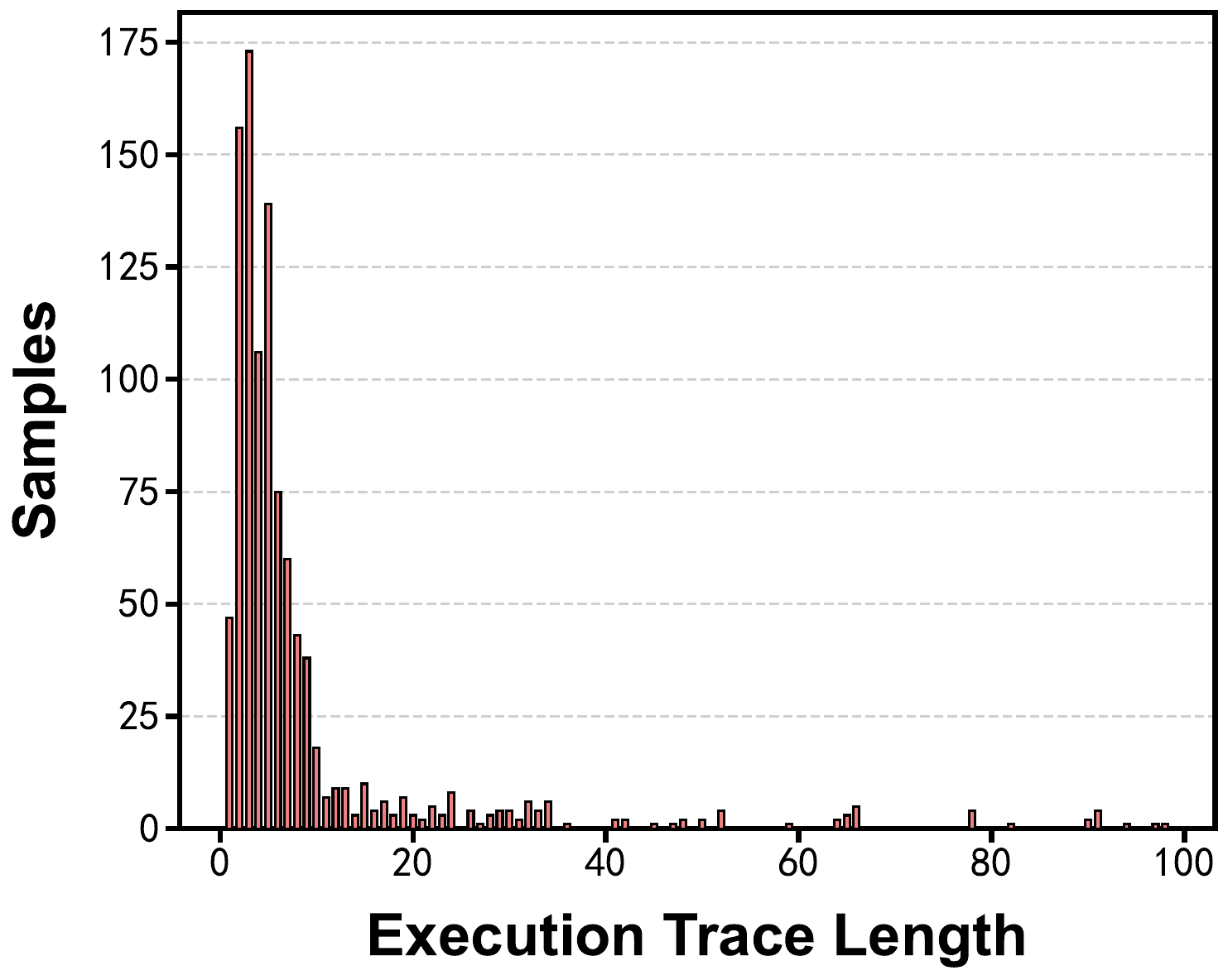}
    \caption*{(a) Distribution}
  \end{minipage}
  \hfill
  \begin{minipage}[b]{0.49\linewidth}
    \centering
    \includegraphics[width=\linewidth]{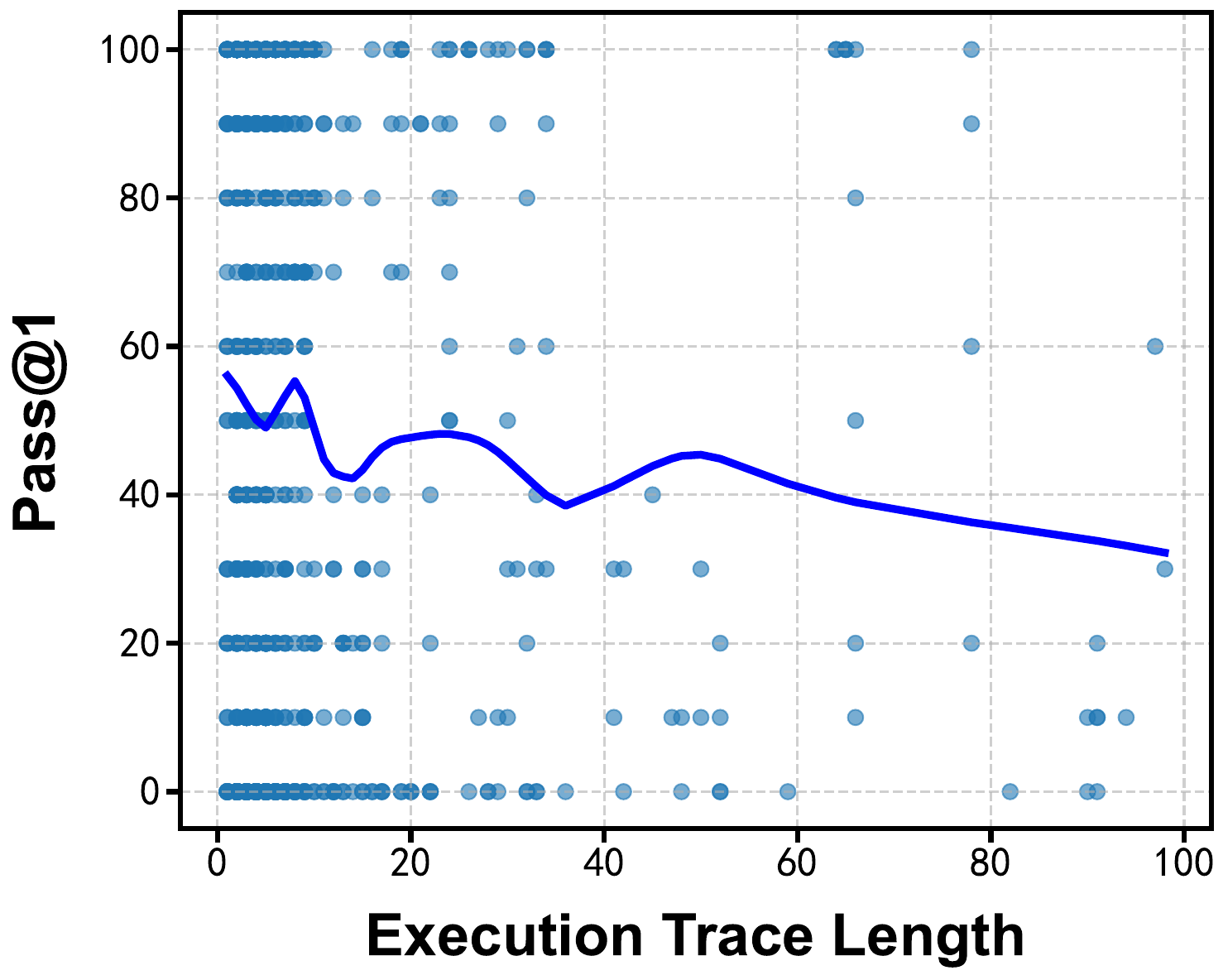}
    \caption*{(b) Performance impact}
  \end{minipage}
  \label{fig:ds1000_exec}
  \vspace{-1em}
\end{figure}

We analyze six features on DS-1000 to understand API reasoning challenges, grouping them by: static composition features (API count, code lines), parameter complexity (defined and specified parameters), and dynamic execution features (execution trace length) as shown in Figures~\ref{fig:ds1000_apicount}-\ref{fig:ds1000_exec}.

\textbf{API composition complexity} exhibits a generally declining trend with increasing API count. Performance decreases steadily from single-API snippets (~70\%) to those with 4--5 APIs (~40\%), after which it largely stabilizes. This suggests that the marginal difficulty of adding APIs diminishes once the composition reaches a certain level of complexity, possibly because highly complex snippets follow more formulaic API chaining patterns. 
Code lines show a weaker impact compared to CRUXEval, with performance remaining relatively stable for snippets under 10 lines before declining more noticeably for longer code, confirming that high-level API abstractions shift the primary burden from syntax parsing to semantic understanding.

\textbf{Parameter complexity} emerges as a notable challenge in API scenarios. Defined parameters (total parameters across API signatures) show a sharp initial decline---from approximately 80\% pass@1 at low counts to around 50\% at moderate values---though the relationship becomes noisier at higher counts where data is sparser. 
Specified parameters (explicitly provided non-default arguments) exhibit a more consistent and pronounced negative correlation, with performance declining steadily from ~78\% to below 30\% as the number of specified parameters increases. 
This contrast suggests that explicitly configured parameters, which require understanding of how specific argument values alter API behavior, pose a greater reasoning challenge than merely recognizing API signatures with their default configurations.

\textbf{Execution trace length} again shows a clear negative association with performance, though the trend is less uniform than in the intrinsic logic scenario. Despite shorter visible code, DS-1000 snippets contain substantial execution traces due to complex API internal operations. However, the relationship exhibits more variability in the mid-range, possibly reflecting the heterogeneous nature of API execution patterns across different libraries. This highlights that even with high-level abstractions, the underlying execution complexity---hidden within API implementations---remains an important factor in code reasoning difficulty.

These findings point to several practical considerations. The steady decline in performance with API count, suggests that compositional reasoning strategies could benefit from hierarchically decomposing multi-API interactions into smaller reasoning units. The particularly strong impact of specified parameters indicates that parameter-level semantics---how specific argument configurations affect API behavior---represent a distinct and underexplored dimension of reasoning difficulty beyond function-level understanding. More broadly, the interplay between visible code structure and hidden execution complexity underscores the need for augmentation approaches that help models reason about encapsulated API logic, not just the surface-level code they can observe.

\subsubsection{Scenario 3: Unseen Function usage reasoning}

\begin{figure}[htbp]
  \caption{Unseen API count analysis in MonkBeatEval.}
  \Description{Two-panel figure for MonkBeatEval unseen API count.}
  \centering
  \begin{minipage}[b]{0.49\linewidth}
    \centering
    \includegraphics[width=\linewidth]{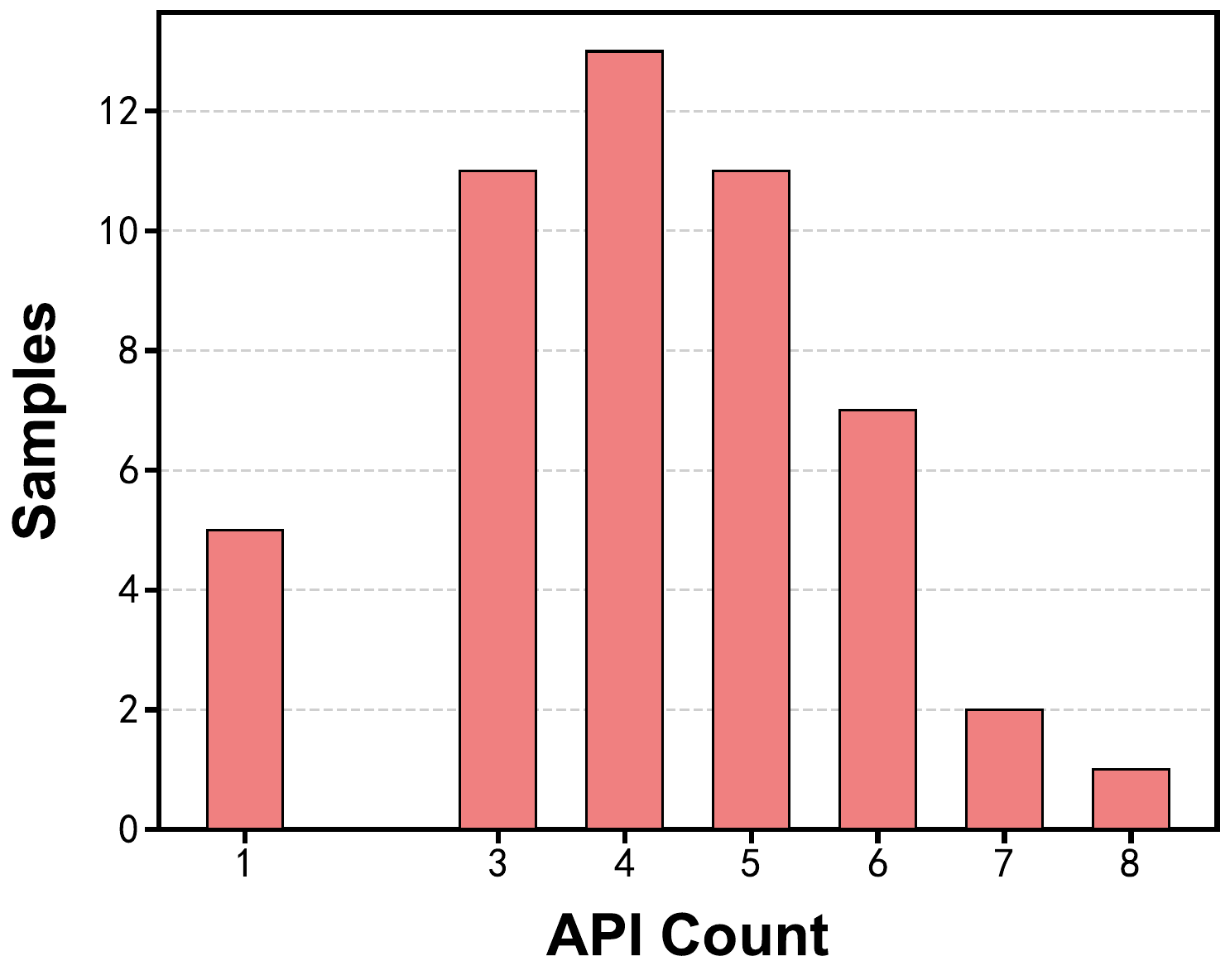}
    \caption*{(a) Distribution}
  \end{minipage}
  \hfill
  \begin{minipage}[b]{0.49\linewidth}
    \centering
    \includegraphics[width=\linewidth]{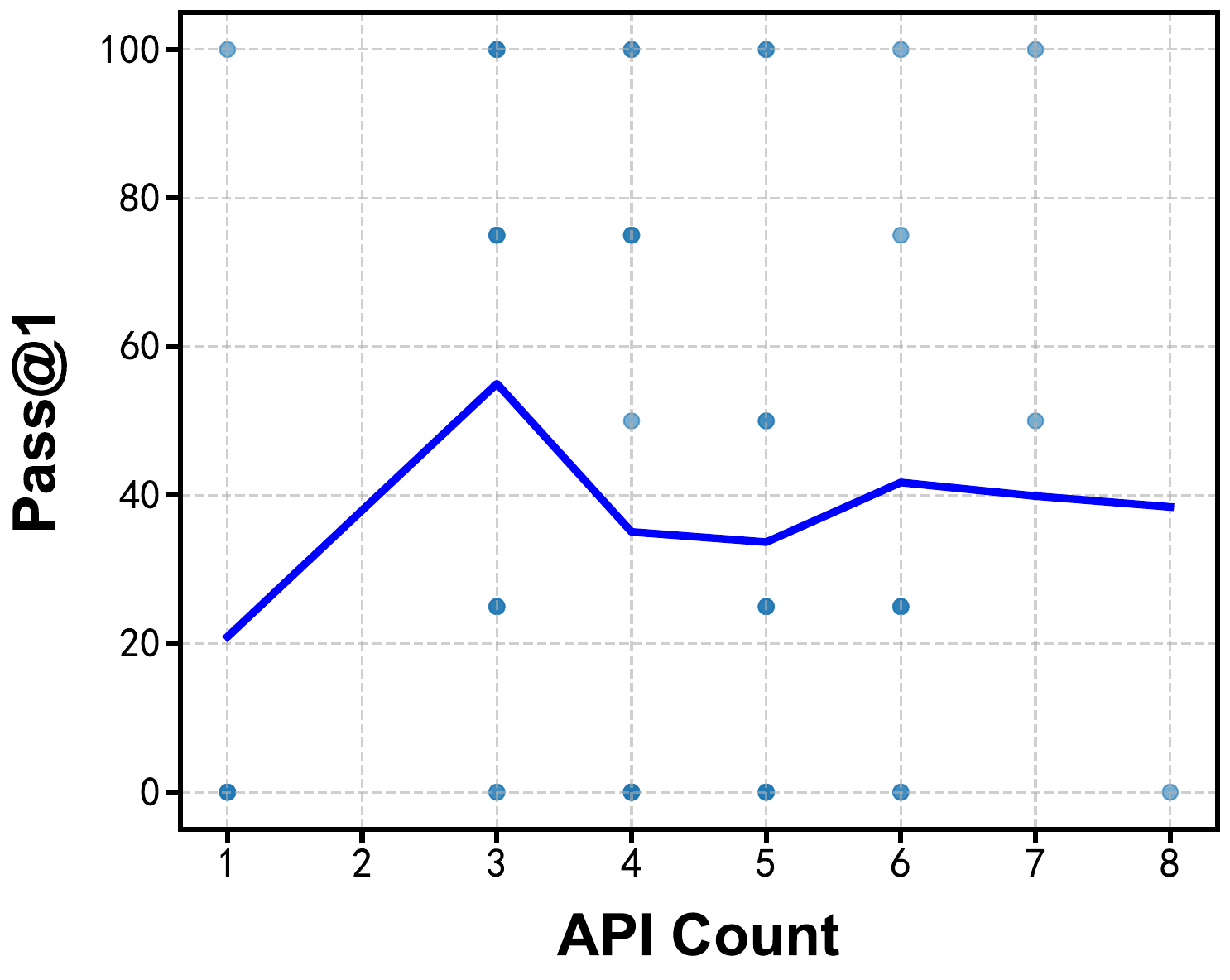}
    \caption*{(b) Performance impact}
  \end{minipage}
  \label{fig:monkbeat_apicount}
\end{figure}

We analyze unseen API count on MonkBeatEval, which contains novel functions outside the model's training distribution (Figure~\ref{fig:monkbeat_apicount}).

The unseen function scenario reveals a distinctive pattern compared to familiar API reasoning. Beyond 3 unseen APIs, performance generally declines from a peak of approximately 55\% to 35--40\% in higher counts, suggesting that the cumulative uncertainty from multiple unknown functions increasingly challenges models as they must simultaneously infer each function's semantics while tracking their compositional interactions. Interestingly, performance at the lowest API count (1 unseen API) is also relatively low (~20\%), resulting in a non-monotonic overall pattern. One possible explanation is that snippets involving multiple unseen APIs provide richer contextual signals---such as coordinated variable naming, data flow patterns---that help models infer function behavior through mutual context, whereas single-API snippets offer fewer surrounding cues. Given the relatively small dataset size (50 samples).
Compared with the familiar API scenario in DS-1000, where performance declines steadily from single-API snippets onward, this pattern highlights a qualitative difference: for known APIs, each additional call purely adds compositional complexity, whereas for unseen functions, additional calls simultaneously introduce uncertainty and provide inferential context---two competing forces whose balance shifts as the number of unknown functions grows.

\paragraph{Cross-Scenario Synthesis.}
Comparing feature impacts across the three scenarios reveals several recurring patterns. 
First, \textbf{dynamic execution features}---including execution trace length and runtime parameters---consistently show stronger associations with performance degradation than \textbf{static features} such as code lines and cyclomatic complexity. In both the intrinsic logic and API interaction scenarios, execution trace length emerges as a key difficulty driver, indicating that simulating multi-step runtime behavior poses a more fundamental challenge for current LLMs than parsing static code structure. 
Second, \textbf{API composition complexity} shows declining performance trends as the number of interacting components increases, though the specific patterns differ by context: a steady decline for familiar APIs in DS-1000 and a non-monotonic pattern for unseen functions in MonkBeatEval, where a moderate number of co-occurring unknown function calls may provide mutual context for reasoning. 
Third, the parameter-level analysis in DS-1000 reveals that \textbf{specified parameters} have a particularly pronounced impact, suggesting that understanding how specific argument values alter API behavior represents a distinct reasoning challenge beyond recognizing API signatures. 
These findings collectively suggest that advancing code reasoning capabilities may benefit more from improving models' ability to simulate execution dynamics and reason about runtime behavior than from enhancing static code comprehension alone.

\begin{mybox}
  \small
  \textbf{Answer to RQ2:}
  Dynamic execution features consistently impact reasoning more than static code structure across scenarios. API composition complexity and parameter-level semantics emerge as key difficulty drivers, with specified parameters showing particularly pronounced effects. For unseen functions, the relationship between API count and performance differs qualitatively from familiar API scenarios, reflecting the tension between contextual inference and compounding uncertainty. These findings suggest that advancing code reasoning may benefit more from improving execution simulation and runtime semantic understanding than from enhancing static code comprehension alone.
\end{mybox}

\subsection{RQ3: How effective are common reasoning strategies across different code reasoning scenarios?}

Enhancing LLMs' code understanding capabilities has been a central focus in software engineering research. Two prominent approaches have emerged: \textbf{Chain-of-Thought (CoT)} prompting, which elicits step-by-step reasoning to decompose complex problems~\cite{22cot}, and \textbf{Retrieval-Augmented Generation (RAG)}, which augments LLM knowledge by retrieving relevant external information~\cite{20rag}. While these strategies have demonstrated substantial improvements in code generation tasks, their effectiveness for code reasoning---particularly across scenarios with varying knowledge requirements---remains underexplored.

Note that we focus on these lightweight augmentation strategies rather than agentic workflows (e.g., ReAct loops~\cite{23react}), since CodeGlance targets efficient code comprehension scenarios where developers need immediate understanding at a glance, and iterative multi-turn approaches incur prohibitive latency overhead for such real-time assistance. In this RQ, we systematically evaluate how CoT and RAG impact code reasoning performance across our three scenarios, providing actionable insights for practitioners.

\subsubsection{Scenario 1: Intrinsic Logic Reasoning}

\begin{table}[t]
\centering
\caption{Impact of reasoning strategies on CRUXEval.}
\label{tab:cruxeval_enhance}
\begin{tabular}{llcc}
\toprule
\multirow{2}{*}{\textbf{Model}} & \multirow{2}{*}{\textbf{Method}} & \multicolumn{2}{c}{\textbf{CRUXEval}} \\
\cmidrule(lr){3-4}
 &  & \textbf{pass@1} & \textbf{pass@3} \\
\midrule
\multirow{2}{*}{Qwen2.5-3b} & Direct & 21.5\% & 38.3\% \\
& CoT fewshot & 46.1\% & 63.8\% \\
\midrule
\multirow{2}{*}{Qwen2.5-7b} & Direct & 46.2\% & 56.8\% \\
& CoT fewshot & 56.2\% & 72.2\% \\
\midrule
\multirow{2}{*}{Qwen2.5-14b} & Direct & 58.4\% & 59.0\% \\
& CoT fewshot & 74.8\% & 78.3\% \\
\midrule
\multirow{2}{*}{Qwen2.5-32b} & Direct & 65.3\% & 68.5\% \\
& CoT fewshot & 77.9\% & 85.1\% \\
\bottomrule
\end{tabular}
\end{table}

For intrinsic logic reasoning on CRUXEval (Table~\ref{tab:cruxeval_enhance}), CoT few-shot prompting yields consistent improvements across all model scales, with gains inversely related to model size: +24.6\% for the 3B model (21.5\% → 46.1\%) compared to +12.6\% for the 32B model (65.3\% → 77.9\%). Notably, the 3B model with CoT few-shot nearly matches the 7B direct baseline (46.1\% vs.\ 46.2\%), suggesting that enforced step-by-step reasoning can substantially compensate for limited model capacity in logic-intensive tasks. 
\subsubsection{Scenario 2: API Interaction Reasoning}

\begin{table}[t]
\centering
\caption{Impact of reasoning strategies on DS1000-R.}
\label{tab:ds1000_enhance}
\begin{tabular}{llcc}
\toprule
\multirow{2}{*}{\textbf{Model}} & \multirow{2}{*}{\textbf{Method}} & \multicolumn{2}{c}{\textbf{DS1000-R}} \\
\cmidrule(lr){3-4}
 &  & \textbf{pass@1} & \textbf{pass@3} \\
\midrule
\multirow{4}{*}{Qwen2.5-3b} & Direct & 37.5\% & 54.7\% \\
& CoT zeroshot & 29.8\% & 51.7\% \\
& CoT fewshot & 37.9\% & 62.1\% \\
& RAG (API doc) & 18.0\% & 37.2\% \\
\midrule
\multirow{4}{*}{Qwen2.5-7b} & Direct & 49.9\% & 63.0\% \\
& CoT zeroshot & 50.0\% & 63.8\% \\
& CoT fewshot & 58.7\% & 77.1\% \\
& RAG (API doc) & 50.5\% & 64.2\% \\
\midrule
\multirow{4}{*}{Qwen2.5-14b} & Direct & 62.0\% & 63.0\% \\
& CoT zeroshot & 61.1\% & 61.6\% \\
& CoT fewshot & 63.5\% & 75.6\% \\
& RAG (API doc) & 62.4\% & 63.1\% \\
\midrule
\multirow{4}{*}{Qwen2.5-32b} & Direct & 66.2\% & 69.7\% \\
& CoT zeroshot & 61.6\% & 78.1\% \\
& CoT fewshot & 68.0\% & 82.2\% \\
& RAG (API doc) & 67.7\% & 80.5\% \\
\bottomrule
\end{tabular}
\end{table}

For API interaction reasoning on DS-1000 (Table~\ref{tab:ds1000_enhance}), CoT few-shot consistently outperforms other strategies, improving pass@1 by up to +8.8\% for 7B models and pass@3 by +12.5\% for 32B models. CoT zero-shot shows more mixed results: it degrades pass@1 for both the 3B model (37.5\%→29.8\%) and the 32B model (66.2\%→61.6\%), yet improves pass@3 for the 32B model (69.7\%→78.1\%), suggesting that without structured examples, self-guided reasoning introduces variability that only benefits larger models when multiple attempts are allowed.

RAG with API documentation shows a striking capacity dependence: it severely degrades the 3B model (37.5\%→18.0\% pass@1), while providing modest gains for the 32B model (pass@3: 69.7\%→80.5\%). This suggests that when LLMs already possess substantial API knowledge from pretraining, smaller models lack the capacity to effectively filter and leverage redundant documentation, whereas larger models can selectively extract useful information from the augmented context.

\subsubsection{Scenario 3: Unseen Function Reasoning}

\begin{table}[t]
\centering
\caption{Impact of reasoning strategies on PanNumEval-R/MonkBeatEval-R.}
\label{tab:monkbeat_enhance}
\resizebox{\columnwidth}{!}{
\begin{tabular}{llcccc}
\toprule
\multirow{2}{*}{\textbf{Model}} & \multirow{2}{*}{\textbf{Method}} & \multicolumn{2}{c}{\textbf{PanNumEval-R}} & \multicolumn{2}{c}{\textbf{MonkBeatEval-R}} \\
\cmidrule(lr){3-4} \cmidrule(lr){5-6}
 &  & \textbf{pass@1} & \textbf{pass@3} & \textbf{pass@1} & \textbf{pass@3} \\
\midrule
\multirow{5}{*}{Qwen2.5-3b} & Direct & 15.0\% & 33.0\% & 6.0\% & 14.5\% \\
& CoT fewshot & 34.0\% & 60.0\% & 26.0\% & 48.5\% \\
& RAG (API doc) & 19.5\% & 39.5\% & 10.0\% & 22.0\% \\
& Code Search & 19.0\% & 40.0\% & 14.0\% & 32.5\% \\
\midrule
\multirow{5}{*}{Qwen2.5-7b} & Direct & 52.0\% & 63.0\% & 40.5\% & 57.5\% \\
& CoT fewshot & 54.5\% & 82.0\% & 43.5\% & 64.0\% \\
& RAG (API doc) & 54.0\% & 68.0\% & 42.0\% & 61.0\% \\
& Code Search & 49.0\% & 63.5\% & 35.0\% & 55.5\% \\
\midrule
\multirow{5}{*}{Qwen2.5-14b} & Direct & 56.0\% & 56.0\% & 49.5\% & 50.0\% \\
& CoT fewshot & 77.5\% & 83.5\% & 71.0\% & 77.0\% \\
& RAG (API doc) & 57.5\% & 58.0\% & 51.5\% & 53.5\% \\
& Code Search & 55.5\% & 56.0\% & 58.0\% & 58.0\% \\
\midrule
\multirow{5}{*}{Qwen2.5-32b} & Direct & 67.0\% & 73.0\% & 64.0\% & 79.5\% \\
& CoT fewshot & 74.0\% & 89.5\% & 61.0\% & 80.5\% \\
& RAG (API doc) & 73.5\% & 85.5\% & 71.0\% & 87.0\% \\
& Code Search & 61.5\% & 82.0\% & 64.5\% & 85.5\% \\
\bottomrule
\end{tabular}
}
\end{table}

The unseen function scenario (Table~\ref{tab:monkbeat_enhance}) evaluates whether common enhancement strategies can help LLMs reason about novel functions outside their training distribution. Since MonkBeatEval is derived from PanNumEval by systematically replacing familiar Pandas/NumPy APIs with equivalent but unseen functions imported from Monkey/BeatNum modules, comparing strategy effectiveness across the two datasets allows us to isolate the impact of knowledge gaps from other confounding factors.

CoT few-shot yields substantial improvements on both datasets, with smaller models benefiting most (3B on MonkBeat: 6.0\%→26.0\%; 3B on PanNum: 15.0\%→34.0\%). Yet even with CoT, models consistently score lower on MonkBeatEval than on PanNumEval at the same scale (e.g., 14B: 71.0\% vs.\ 77.5\%), indicating that step-by-step reasoning alone cannot fully compensate for missing API knowledge.

RAG with API documentation provides moderate improvements for unseen function reasoning, though its effectiveness depends heavily on model capacity. Smaller models gain little from retrieved documentation (7B: +1.5\%; 14B: +2.0\%), while the 32B model achieves more substantial improvements (+7.0\% pass@1, +7.5\% pass@3), suggesting that effectively leveraging external documentation during reasoning requires sufficient model capacity to filter and integrate the additional context. Notably, even with RAG augmentation, the performance gap between MonkBeatEval and PanNumEval persists across all scales, indicating that retrieved documentation helps but cannot fully close the disadvantage introduced by unseen functions.

Code search, which retrieves raw source code of unseen functions, shows limited effectiveness compared to API documentation. Natural language documentation consistently outperforms source code retrieval (e.g., 7B on MonkBeat: 42.0\% vs.\ 35.0\%), suggesting that concise semantic descriptions are more readily integrated into the reasoning process than source code, which effectively requires nested code comprehension as a prerequisite.

\paragraph{Cross-Scenario Synthesis.} Three observations emerge from comparing strategy effectiveness across scenarios: (1) \textbf{CoT consistently improves reasoning where knowledge is sufficient}---it yields substantial gains in ILR and AIR scenarios and remains the most effective single strategy across all settings, with smaller models benefiting disproportionately; (2) \textbf{RAG effectiveness depends on both knowledge availability and model capacity}---it provides moderate gains for unseen function reasoning but requires sufficient model scale to be leveraged effectively; (3) \textbf{Strategy selection should match the performance bottleneck}---reasoning augmentation for logical complexity, knowledge augmentation for genuine knowledge gaps, with model capacity as a prerequisite for both.

\begin{mybox}
  \small
  \textbf{Answer to RQ3:}
  CoT few-shot is the most broadly effective strategy, improving reasoning across all scenarios with smaller models benefiting disproportionately. RAG with API documentation provides moderate gains for unseen function reasoning but depends heavily on model capacity, and cannot fully bridge the performance gap caused by knowledge deficits. Strategy effectiveness depends critically on matching the bottleneck: reasoning augmentation for logical complexity, knowledge augmentation for novelty---with sufficient model capacity as a prerequisite for leveraging either effectively.
\end{mybox}
\section{Discussion and Threats to Validity}
\paragraph{Benchmark Representativeness}
Our benchmark focuses on three specific scenarios (intrinsic logic, API interaction, and unseen functions) which may not capture all real-world code reasoning contexts. To ensure representativeness, we deliberately selected scenarios based on systematic analysis of common developer workflows---code review (understanding unfamiliar logic), API integration (composing library functions), and repository navigation (reasoning about custom utilities). We leveraged established datasets (CRUXEval, DS-1000, TorchdataCode) that have been validated in prior work and constructed MonkBeatEval following rigorous benchmark design principles. The diversity of our three scenarios, spanning different reasoning challenges and knowledge requirements, provides comprehensive coverage of core code understanding tasks encountered in practice.

\paragraph{Model Selection and Generalizability}
Our evaluation covers 7 state-of-the-art LLMs, which may not represent all available models. However, we strategically selected models spanning different architectures (GPT-4o, Qwen2.5-Coder, DeepSeek-Coder), scales (3B-32B parameters for open-source models), and training paradigms (closed-source vs. open-source, general-purpose vs. code-specialized). This diversity ensures our findings reflect broader trends rather than model-specific artifacts. The consistent patterns observed across different model families (e.g., dynamic features outweighing static features) strengthen the generalizability of our conclusions. While future models may exhibit different absolute performance levels, the relative challenges and reasoning patterns we identify are likely to persist.

\paragraph{Feature Coverage}
We analyze 9 code complexity features across three categories (logic, API, knowledge), which may not exhaustively capture all factors affecting code reasoning. Our feature selection was guided by both theoretical considerations (prior work on code complexity metrics) and practical relevance (features developers commonly encounter). We prioritize features with clear operational definitions and reliable automatic extraction to ensure reproducibility. The strong correlations observed between our selected features and reasoning performance (particularly dynamic execution features) validate their relevance. While additional features (e.g., variable naming quality, comment density) could provide further insights, our current feature set identifies the primary bottlenecks---dynamic execution simulation and compositional complexity---that should be prioritized for improving code reasoning systems.

\section{Related Work}
\paragraph{Code Understanding Benchmarks}
Evaluating LLMs on code has evolved from basic generation tasks---HumanEval~\cite{21humaneval}, MBPP~\cite{mbpp}, CodeSearchNet~\cite{20codesearchnet}, CodeXGLUE~\cite{21codexglue}---to complex reasoning scenarios including translation~\cite{codetrans}, synthesis~\cite{codex}, bug detection~\cite{bugdetection}, and class-level generation~\cite{23classeval}. Recent benchmarks emphasize holistic evaluation: LiveCodeBench~\cite{25livecodebench} addresses contamination with time-evolving problems, BigCodeBench~\cite{25bigcodebench} evaluates diverse function calls across 43 languages, CodeScope~\cite{24codescope} assesses multilingual multitask performance, and SWE-bench~\cite{swebench} tests real-world issue resolution. For code reasoning specifically, CRUXEval~\cite{cruxeval} introduced input-output prediction, DS-1000~\cite{ds1000} focused on data science APIs, CodeMind~\cite{24CodeMind} challenged reasoning robustness, NExT~\cite{24NExT} taught execution reasoning, and CodeReason~\cite{25CodeReason} examined runtime behavior understanding. However, existing benchmarks evaluate isolated scenarios without systematically investigating how code characteristics (execution complexity, API composition, knowledge familiarity) affect reasoning. CodeGlance addresses this gap through three orthogonal scenarios with targeted feature analysis, revealing what fundamentally challenges LLM code reasoning.

\paragraph{LLM Code Understanding and Enhancement Strategies}
Code-specialized LLMs---Codex~\cite{codex}, CodeLlama~\cite{codellama}, StarCoder~\cite{starcoder}, DeepSeek-Coder~\cite{deepseek-v3}, Qwen-Coder~\cite{qwen25}---have achieved remarkable performance through targeted pre-training. Enhancement strategies fall into three categories. \textit{Reasoning augmentation} employs Chain-of-Thought prompting~\cite{22cot} for step-by-step decomposition in generation~\cite{cotcode} and repair~\cite{cotapr, 23Self-Edit, 24Self-Debug}, ReAct~\cite{23react} for iterative reasoning-action cycles, and process-based feedback~\cite{24processreward} for execution-aware refinement. \textit{Knowledge augmentation} leverages Retrieval-Augmented Generation~\cite{20rag, 23repocoder} for API documentation~\cite{ragcode, 24compositionalapi, 25exploracoder}, code search for contextual suggestions~\cite{24Search-APR2}, and repository-level context~\cite{autocoderover}. \textit{Debugging and repair systems} integrate dynamic analysis~\cite{24LDB, 25AutoSD} with agentic workflows~\cite{25RepairAgent, 24FixAgent, 24sweagent, 25lingmagpt} for autonomous program improvement. Despite these advances, their effectiveness across reasoning scenarios with varying execution complexity and knowledge requirements remains unclear. Our systematic evaluation reveals scenario-specific strategy efficacy: CoT consistently improves reasoning across scenarios, RAG provides capacity-dependent gains primarily for knowledge-gap scenarios, yet neither fully bridges the gap for unseen functions, exposing fundamental limitations in compositional generalization.

\section{Conclusion}
We present CodeGlance, a multi-dimensional benchmark evaluating LLMs' code reasoning across three realistic scenarios. Our evaluation of 7 LLMs reveals that unseen functions pose significant challenge, dynamic features outweigh static structure, and augmentation strategies must match challenge types. These findings guide development of more capable code reasoning systems.

\bibliographystyle{ACM-Reference-Format}
\bibliography{main}

\appendix

\end{document}